\documentclass{article}
\usepackage{authblk}
\usepackage[utf8]{inputenc}
\usepackage[english]{babel}
\usepackage[nottoc]{tocbibind}
\usepackage{csquotes}
\usepackage[backend=biber, style=apa]{biblatex}
\usepackage{libertine}
\usepackage[libertine]{newtxmath}

\usepackage{tgcursor}
\usepackage{booktabs}
\usepackage{siunitx}
\usepackage[a4paper, margin=1in]{geometry}
\usepackage{amsmath}
\usepackage{amssymb}
\usepackage{mathtools}
\usepackage[shortlabels]{enumitem}
\usepackage{hyperref}
\usepackage{float}
\usepackage{ccaption}
\usepackage[usenames,dvipsnames]{xcolor}
\usepackage{graphicx}
\usepackage{subcaption}
\usepackage{mwe}
\usepackage{listings}
\usepackage{tikz}
\usepackage{algorithm}
\usepackage{algpseudocode}
\usepackage{eqparbox,array}
\usepackage{xspace}
\usepackage{comment}
\usepackage[noabbrev,capitalize]{cleveref}
\creflabelformat{equation}{#2\textup{#1}#3}

\numberwithin{equation}{section}
\usepackage{bbm}
      
\newcommand{\dd}{\mathrm{d}}

\newcommand{\vct}[1]{\mathbf{#1}} 
\newcommand{\mat}[1]{\mathbf{#1}}

\graphicspath{ {./} }

\AtEveryBibitem{%
  \clearfield{note}%
  \clearfield{url}%
  \clearfield{urlyear}%
  \clearfield{issn}%
  \clearlist{language}
}
\makeatletter
\DeclareRobustCommand\onedot{\futurelet\@let@token\@onedot}
\def\@onedot{\ifx\@let@token.\else.\null\fi\xspace}

\makeatother

\title{Predictive Coding Theories of Cortical Function}
\author[1, 2, 3]{Linxing Preston Jiang}
\author[1, 2, 3]{Rajesh P. N. Rao\thanks{Corresponding Author}}
\affil[1]{Paul G. Allen School of Computer Science \& Engineering, University of Washington}
\affil[2]{Center for Neurotechnology, University of Washington}
\affil[3]{Computational Neuroscience Center, University of Washington}
\affil[ ]{\texttt {\small \{prestonj,rao\}@cs.washington.edu}}
\date{}

\AtBeginDocument{%
}

\allowdisplaybreaks

\begin{document}

\begin{center}
    Cite as: Jiang, L. P., \& Rao, R. P. N. (2022). Predictive coding theories of cortical function. In Oxford Research Encyclopedia of Neuroscience. doi: \url{https://doi.org/10.1093/acrefore/9780190264086.013.328}
\end{center}
{\let\newpage\relax\maketitle}

\begin{abstract}
Predictive coding is a unifying framework for understanding perception, action and neocortical organization. In predictive coding, different areas of the neocortex implement a hierarchical generative model of the world that is learned from sensory inputs. Cortical circuits are hypothesized to perform Bayesian inference based on this generative model. Specifically, the Rao-Ballard hierarchical predictive coding model assumes that the top-down feedback connections from higher to lower order cortical areas convey predictions of lower-level activities. The bottom-up, feedforward connections in turn convey the errors between top-down predictions and actual activities. These errors are used to correct current estimates of the state of the world and generate new predictions. Through the objective of minimizing prediction errors, predictive coding provides a functional explanation for a wide range of neural responses and many aspects of brain organization.
\end{abstract}

\textbf{\textit{Subjects}}: Computational Neuroscience

\tableofcontents

\section{Introduction}

\begin{figure}
    \centering
    \includegraphics{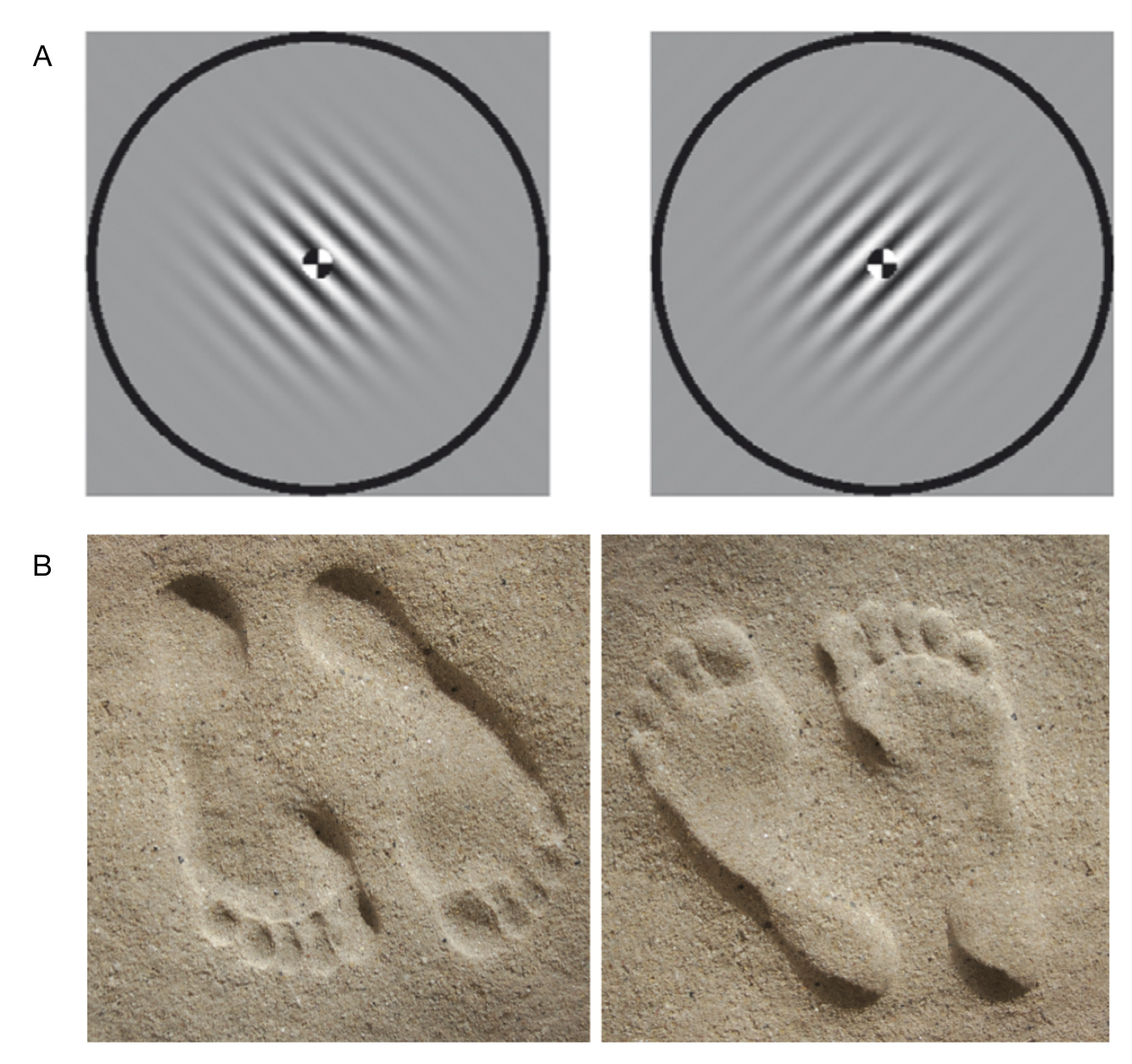}
    \caption{\textbf{Bayesian perception}. \textbf{(A)} Example stimuli (dichoptic orthogonal gratings) presented to the left and right eye simultaneously that induce binocular rivalry. Subjects perceive exclusively one of the two grating orientations, with perception switching between the two orientations every few seconds. Adapted from \textcite{tong_neural_2006}. \textbf{(B)} An illustration of the effects of the light-from-above prior assumption that the brain appears to use in 3D visual perception of a 2D image. The image of the two footprints on the right is a 180 degree rotation of the image on the left, but the perception of the shape of the footprints is markedly different. Adapted from \textcite{ernst_multisensory_2011}.}
    \label{fig:bino}
\end{figure}

A normative theory for understanding perception is that the brain uses an internal model of the external world to infer the hidden causes of its sensory inputs and maintain beliefs about these causes. In the early work of \textcite{gregory_perceptions_1980}, perception was defined as hypothesis testing, emphasizing the process of inferring explanations for sensory inputs. The notion that perception is an inference process based on internal models (rather than a purely bottom-up feature-extracting process) is well exemplified by the phenomenon of binocular rivalry \parencite{tong_neural_2006}. Binocular rivalry occurs when conflicting monocular images are presented separately to each of the two eyes (\cref{fig:bino}A) Instead of perceiving a stable mixture or superposition of the two stimuli, the subject perceives exclusively the object or feature in one of the two distinct images presented to each eye, with perception alternating between the two images every few seconds. Such “rivalry” challenges the traditional stimulus-driven feature-extraction view of perception – why would perception alternate between two interpretations if the process is completely bottom-up, given that the stimulus does not change? When perception is viewed as forming hypotheses to infer the hidden causes of images, binocular rivalry can be understood as the brain entertaining two competing hypotheses to explain a conflicting sensory input. 

Having an internal model of the environment also helps disambiguate sensory inputs with multiple interpretations. \cref{fig:bino}B shows an example: the two footprints on the right appear to be convex (oriented upward towards the viewer) while the two on the left appear to be concave (oriented downward away from the viewer). However, the image on the right is the same as the image on the left, only rotated 180 degrees. The two different interpretations of the footprints arise from the brain using a “light-from-above” prior assumption \parencite{sun_where_1998}: the brain’s internal model assumes that light sources tend to be above the observer, an ecologically valid assumption. Such assumptions are necessary because visual perception is an ill-posed problem: multiple 3D configurations can give rise to the same 2D image due to the projection of the 3D world onto a 2D retina, making assumptions such as “light-from-above" necessary for inferring properties of visual objects. Note that the observer is typically not aware of such prior assumptions but rather, they are incorporated by the neural circuits subconsciously to compute beliefs over hidden causes through the dynamics of neural activities (thereby implementing perception as “unconscious inference”). As part of the internal model, such priors can be expected to be adapted to the environment that the organism lives in.

How can neural circuits in the cortex learn internal models of the world, and how can such circuits combine prior beliefs with sensory evidence for Bayesian inference? Predictive coding offers a possible neural implementation. The predictive coding model of \textcite{rao_predictive_1999} assumes that the areas comprising the cortical hierarchy \parencite{hubel_receptive_1959, felleman_distributed_1991} implement a hierarchical generative model of the sensory world. The neural activities at each level of the hierarchy represent the brain’s internal belief of the hidden causes of the stimuli at a particular abstraction level (e.g., edges, object parts, objects). Furthermore, the model assumes that the top-down feedback connections from higher to lower order cortical areas convey predictions of lower-level activities. The bottom-up feedforward connections in turn convey prediction errors, calculated as the difference between the top-down predictions and actual activities. The neural activities at each level representing the beliefs about the hidden causes are jointly influenced by both the top-down predictions and the bottom-up error signals. Overall, the model assumes the goal of the cortex is to minimize prediction errors across all levels. Importantly, the above neural operations can be interpreted within a Bayesian framework: the top-down predictions convey prior beliefs based on learned expectations while the bottom-up prediction errors carry evidence from the current input. Predictive coding combines these two sources of information, weighted according to their reliability (inverse variances or “precisions”), to compute the posterior beliefs over hidden causes at each level. The objective of minimizing prediction errors across all levels can thus be shown to be equivalent to finding the \textit{maximum a posteriori} (MAP) estimates of the hidden causes. 

The phrase “predictive coding” was originally used to capture a form of efficient coding. The center-surround receptive fields and biphasic temporal antagonism in responses of cells in the retina and lateral geniculate nucleus (LGN) can be interpreted as performing decorrelation through a simple form of predictive coding: rather than conveying the local intensity directly, retinal and LGN cells can be interpreted as sending the differences (errors) between the local intensity and a prediction of that intensity computed as a linear weighted sum of nearby values in space and preceding input values in time \parencite{srinivasan_predictive_1982, dong_statistics_1995, huang_predictive_2011}. In auditory information processing, \textcite{smith_efficient_2006} used the same efficient coding principle to derive a model which yields kernels (filter weights) that closely match auditory filters.

More broadly, predictive coding can be viewed as Bayesian inference in the context of Rao and Ballard’s hierarchical predictive coding model \parencite{rao_dynamic_1997, rao_predictive_1999}. This model was originally proposed to explain extra-classical receptive field effects and contextual modulation. More recent models inspired by predictive coding have demonstrated that a network trained to predict future inputs can explain a number of other cortical properties \parencite{singer_sensory_2018, lotter_neural_2020}. Beyond the cortex, the idea of computing errors between top-down predictions and lower-level inputs is consistent with theories of the cerebellum \parencite{bell_synaptic_1997, wolpert_internal_1998} and models of dopamine responses as reward prediction errors \parencite{schultz_neural_1997}. These examples suggest that the general principle of predictive coding could be a widely applicable and flexible algorithmic strategy implemented by the brain across different regions to support perception, motor control, and reward-based learning.

Empirical evidence for prediction and prediction error signals in the cortex has been growing at a fast pace. Neural responses corresponding to prediction errors induced by visual mismatches during self-generated locomotion have been discovered in layer 2/3 of the primary visual cortex (V1) in rodents \parencite{keller_sensorimotor_2012, fiser_experience-dependent_2016}. Predictive signals have been found in V1 when an animal is adapted to visual-locomotion coupling in a virtual environment \parencite{fiser_experience-dependent_2016}. The cortex also learns to predict novel auditory stimuli coupled to an animal’s locomotion and once learned, suppresses the responses to the learned stimuli in primary auditory cortex \parencite{schneider_cortical_2018}, consistent with prediction error minimization. More recent studies \parencite{jordan_opposing_2020} have found some support for the distinct computational roles of the laminar structure of cortical columns proposed by predictive coding theories. Recent research has also found that unexpected stimuli which induce large prediction error signals can drive synaptic learning in neural circuits \parencite{gillon_learning_2021}, as expected in a predictive coding circuit that uses prediction errors to learn a generative model of the world.

This review is organized as follows. The section “Predictive Coding Models: An Overview” introduces the Rao-Ballard predictive coding model \parencite{rao_predictive_1999} and several related models, as well as the relationship to the free energy principle and active inference. The section “Predictive Coding in the Visual System” discusses the application of hierarchical predictive coding to the visual cortex, explaining classical and extra-classical receptive field effects in V1 in terms of prediction error minimization, followed by a review of experimental studies investigating predictive coding in the neocortex in the section “Empirical Evidence for Predictive Coding”. The final section discusses open questions pertaining to predictive coding and potential future directions.

\section{Predictive Coding Models: An Overview}
\setcounter{section}{2}
The predictive coding model of Rao \& Ballard begins with the assumption that sensory inputs are being generated by hidden states or “causes” in the external world via an unknown generative model. The goal of the brain then is to learn this generative model over many inputs. Perception, for any given sensory input $\vct{I}$, involves inverting this generative model, that is, estimating the hidden states or causes of input $\vct{I}$ given a learned generative model. Neural activities in the predictive coding model are assumed to represent estimates of the hidden state (also known as the latent variable) vector as estimated by the predictive coding neural network, given the observed sensory input vector $\vct{I}$. The prior distribution of hidden states is assumed to be $p(\vct{r})$, which imposes a constraint on neural activities such as sparse activation. The observation model $p(\vct{I} | \vct{r})$ is the likelihood that input $\vct{I}$ is generated given the cause or hidden state $\vct{r}$. The predictive coding model assumes that $p(\vct{I} | \vct{r})$ is parameterized by a matrix $\mat{U}$, which is assumed to be learned
and encoded in the “top-down” synaptic weights of the network. \textit{Inference and learning} correspond respectively to estimating $\vct{r}$ (equivalent to perception) and learning an estimate $\mat{U}$ (corresponding to synaptic learning), both with the goal of maximizing the joint probability $p(\vct{I}, \vct{r})$. Since $p(\vct{I})$ is constant, this is equivalent to maximizing the posterior probability $p(\vct{r} | \vct{I})$, also known as \textit{maximum a posteriori} (MAP) inference.

\subsection{Generative Model of Images}
In the predictive coding model, the likelihood $p(\vct{I} | \vct{r})$ is governed by the following equation, which relates the hidden state $\vct{r}$ to the input $\vct{I}$ via a function $f$ and a matrix $\mat{U}$:
\begin{align}
    \vct{I} = f(\mat{U}\vct{r}) + \vct{n}. \label{eq:img-gen}
\end{align}
Here, $\vct{n}$ is assumed to be zero mean Gaussian noise with covariance $\sigma^2\mathbb{I}$ ($\mathbb{I}$ is the identity matrix). This equation states that the input is assumed to be generated as a linear combination of the columns of matrix $\mat{U}$ weighted by the elements of $\vct{r}$, followed by a function $f$ and additive noise. The function $f$ is a linear or nonlinear function (e.g., identity function, rectification function, or a sigmoidal function). The columns of $\mat{U}$ can be regarded as the “basis” vectors (e.g., edges or “parts” of an image or scene) that can be used to compose an input according to the values in the hidden “causes” vector $\vct{r}$. Given \cref{eq:img-gen} and the fact that $\vct{n}$ is zero mean Gaussian, the negative logarithm of the likelihood $p(\vct{I} | \vct{r})$ can be shown to be proportional to:
\begin{align}
    H_1 = \frac{1}{\sigma^2}\Vert\vct{I} - f(\mat{U}\vct{r})\Vert_2^2 = \frac{1}{\sigma^2}(\vct{I} - f(\mat{U}\vct{r}))^\top(\vct{I} - f(\mat{U}\vct{r})), \label{eq:likelihood}
\end{align}
where $\Vert\vct{x}\Vert_2 = \sqrt{\sum_i x_i^2}$ denotes the Euclidean or $L_2$ norm of vector $\vct{x}$. $H_1$ is the sum of squared errors between the image $\vct{I}$ and its reconstruction (or “prediction”) $f(\vct{Ur})$ across all pixels, weighted by the inverse noise variance (or precision) $\frac{1}{\sigma^2}$. The predictive coding model also allows prior probability distributions $p(\vct{r})$ and $p(\mat{U})$ for the parameters $\vct{r}$ and $\mat{U}$, respectively. Taking these priors into account, we obtain the overall optimization function:
\begin{align}
    H = H_1 + H_2 \label{eq:loss}
\end{align}
with
\begin{align}
    H_2 = g(\vct{r}) + h(\mat{U}), 
\end{align}
where $g(\vct{r})$ and $h(\mat{U})$ are proportional to the negative logarithms of $p(\vct{r})$ and $p(\mat{U})$, respectively. If one assumes that both prior distributions are zero mean Gaussians with inverse variances $\alpha$ and $\lambda$, respectively, one obtains:
\begin{align}
    H_2 = g(\vct{r}) + h(\mat{U}) = \alpha\sum_j r_j^2 + \lambda\sum_{i,j} U_{ij}^2.
\end{align}
Minimizing the overall optimization function $H$ is thus equivalent to MAP estimation. Predictive coding minimizes this objective function using both inference (of $\vct{r}$) and learning (of $\mat{U}$). Inference of $\vct{r}$ is implemented by a recurrent neural network that performs gradient descent on $H$ with respect to $\vct{r}$ for each input. Remarkably, rather than being chosen a priori, the architecture of the predictive coding neural network is predicted from first principles by the gradient descent equations for optimizing $H$ with respect to $\vct{r}$ (see “Network Dynamics and Synaptic Learning” section for details). The matrix $\mat{U}$ is represented by the synaptic weights of the same network and learned through gradient descent on $H$ with respect to $\mat{U}$ across many inputs.

\subsubsection{Sparse Coding as a Special Case of Predictive Coding}
The sparse coding model of \textcite{olshausen_emergence_1996} for learning simple cell-like receptive fields can be regarded as a special case of the predictive coding model described above. In their model, the choice of the likelihood $p(\vct{I} | \vct{r})$ remains the same as above, but the prior $p(\vct{r})$ for the hidden state (causes) $\vct{r}$ is assumed to be a heavy-tailed distribution such as a Laplace distribution. Such a prior encourages sparsity in $\vct{r}$ (majority of the elements of $\vct{r}$ are zero or close to zero). Their model does not explicitly assume any specific prior for the synaptic weights $\mat{U}$. The inference and learning processes are almost identical to those for a single-level predictive coding model (see “Network Dynamics and Synaptic Learning” section). When applied to natural image patches, their model produces localized, orientation-selective receptive fields (columns of $\mat{U}$) similar to those of V1 simple cells, compared to using a Gaussian prior, which produces more global receptive fields. Such a sparseness prior promotes statistical independence in the output and encourages efficiency by selecting only a small subset of features to encode information \parencite{olshausen_emergence_1996, olshausen_sparse_1997, barlow_possible_1961}. The underlying assumption here is that objects in the natural world are composed of a wide variety of features (or parts) but any given object is composed of only a small subset of them. This is consistent with the view that the brain evolved to adopt ecologically useful priors for learning its neural representations in its quest to learn an internal model of the world appropriate for the organism’s ecological niche.

\subsubsection{Hierarchical Predictive Coding}
\begin{figure}
    \centering
    \includegraphics[width=\textwidth]{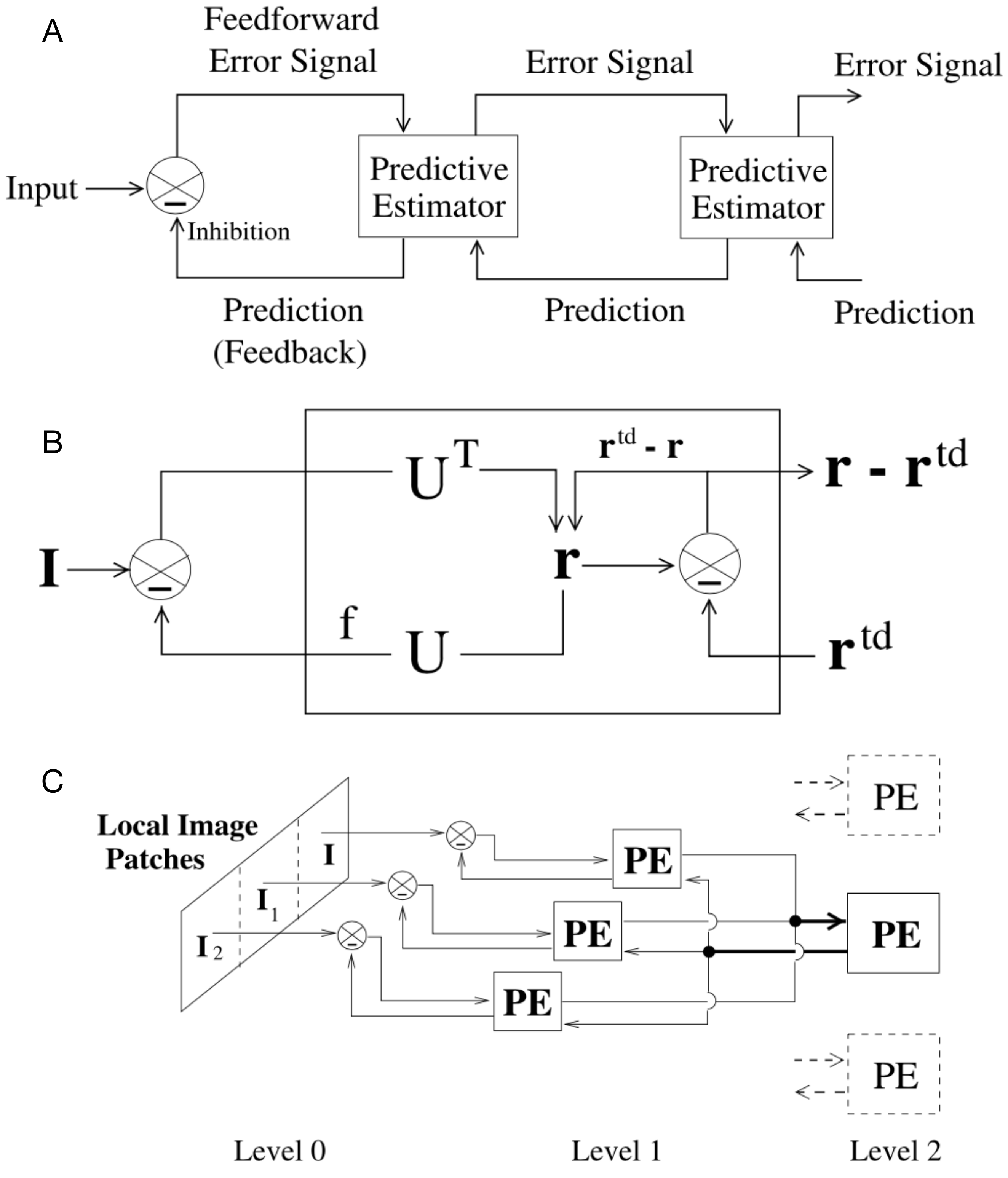}
    \caption{\textbf{Hierarchical Predictive Coding Model.} \textbf{(A)} The general architecture of the hierarchical predictive coding model. Each predictive estimator (PE) module maintains an internal representation, generates top-down prediction (lower arrows, feedback), and receives bottom-up prediction errors (upper arrows, feedforward). \textbf{(B)} Components of a PE module. Feedforward connections carry bottom-up prediction errors from the lower level. Feedback connections deliver top-down predictions to the lower level. The internal representation neurons correct their current estimate $\vct{r}$ using both the bottom-up prediction error and the top-down prediction error. A separate class of error-detecting neurons compute the discrepancy between the current estimate and its top-down prediction, and send the error to the higher level. \textbf{(C)} An example two-level hierarchical network. Three image patches at Level 0 are processed separately by three Level 1 PE modules. These three Level 1 modules converge to provide input (prediction errors) to a single Level 2 module, which attempts to predict the states $\vct{r}$ in all three of these Level 1 PE modules. This convergence effectively increases the receptive field size of neurons as one ascends the hierarchy.}
    \label{fig:hpc}
\end{figure}
The above-described generative model can be extended to multiple hierarchical levels by assuming that the hidden state can be generated by a higher-level representation $\vct{r}^h$, corresponding to more abstract image properties than the lower-level representation:
\begin{align}
    \vct{r} = \vct{r}^{td} + \vct{n}^{td} = f(\mat{U}^h\vct{r}^h) + \vct{n}^{td},
\end{align}
where $\vct{r}^{td} = f(\mat{U}^h\vct{r}^h)$is the top-down prediction of $\vct{r}$ and $\vct{n}^{td}$ is zero mean Gaussian noise with
variance $\sigma_{td}^2$. The lower-level neurons have smaller receptive fields and represent a local image region by estimating the hidden state $\vct{r}$. The higher-level neurons estimate their state $\vct{r}^h$ based on several lower-level hidden states $\vct{r}$ associated with local image patches. This arrangement results in a progressive convergence of inputs from lower to higher levels and an increase in receptive field size as one ascends the hierarchical network, until the receptive fields of the highest-level neurons span the entire input image.

The overall optimization function for the hierarchical predictive coding model is:
\begin{align}
   H=\frac{1}{\sigma^2}(\mathbf{I}-f(\mathbf{U} \mathbf{r}))^{\top}(\mathbf{I}-f(\mathbf{U} \mathbf{r}))+\frac{1}{\sigma_{t d}^2}(\mathbf{r}-\mathbf{r}^{td})^{\top}(\mathbf{r}-\mathbf{r}^{t d})+g(\mathbf{r})+g(\mathbf{r}^h)+h(\mathbf{U})+h(\mathbf{U}^h),
\end{align}
where $g(\vct{r}^h)$ and $h(\mat{U}^{h})$ are terms proportional to the negative logarithm of the priors for $\vct{r}^h$ and $\mat{U}^h$ respectively. Minimizing $H$ is again equivalent to maximizing the posterior $p(\vct{r}, \vct{r}^h, \mat{U}, \mat{U}^h | \vct{I})$. Perceptual inference involves minimizing $H$ with respect to $\vct{r}$ and $\vct{r}^h$ jointly, and learning involves minimizing $H$ with respect to $\mat{U}$ and $\mat{U}^h$. Note that the first level state $\vct{r}$ is now conditioned on the second level state $\vct{r}^h$ and synaptic weights $\mat{U}^h$, but an additional prior constraint such as sparseness may be placed on $\vct{r}$ as well (the $g(\vct{r})$ term). 

\subsubsection{Network Dynamics and Synaptic Learning}
Given the hierarchical generative model above, a MAP estimate of $\vct{r}$ can be obtained using gradient descent on $H$ with respect to $\vct{r}$:
\begin{align}
    \frac{\dd \mathbf{r}}{\dd t}=-\frac{k_1}{2} \frac{\partial H}{\partial \mathbf{r}}=\frac{k_1}{\sigma^2} \mathbf{U}^{\top} \frac{\partial f^{\top}}{\partial \mathbf{x}}(\mathbf{I}-f(\mathbf{U r}))+\frac{k_1}{\sigma_{t d}^2}\left(\mathbf{r}^{t d}-\mathbf{r}\right)-\frac{k_1}{2} g'(\mathbf{r}), \label{eq:dyn}
\end{align}
where $k_1$ is a positive constant governing the rate of descent toward a minimum for $H$, $\vct{x} = \vct{Ur}$, and $g'$ is the derivative of $g$ with respect to $\vct{r}$. A discrete time implementation of the above-mentioned dynamics leads to the following update equation for $\vct{r}$ at each time step (represented by neural activities or firing rates):
\begin{align}
    \hat{\mathbf{r}}_t=\hat{\mathbf{r}}_{t-1}+\frac{k_1}{\sigma^2} \mathbf{U}^{\top} \frac{\partial f^{\top}}{\partial \mathbf{x}_{t-1}}\left(\mathbf{I}-f\left(\mathbf{U} \hat{\mathbf{r}}_{t-1}\right)\right)+\frac{k_1}{\sigma_{t d}^2}\left(\mathbf{r}_{t-1}^{t d}-\hat{\mathbf{r}}_{t-1}\right)-\frac{k_1}{2} g'\left(\hat{\mathbf{r}}_{t-1}\right). \label{eq:dynamics}
\end{align}

This equation, derived from first principles, specifies recurrent network dynamics for hierarchical predictive coding in terms of how the firing rate (or neural response) vector at a given level should be updated over time. At each time step, the neural activity vector $\vct{r}$ is multiplied by the feedback matrix $\mat{U}$ and a new prediction is generated for the lower level (\cref{fig:hpc}A and \cref{fig:hpc}B). This prediction is then subtracted from the lower-level representation $\vct{I}$ to generate the bottom-up error $(\vct{I} - f(\vct{Ur}))$, which is filtered by the feedforward weights $\mat{U}^\top$ and the gradient of the function $f$. Note that the bottom-up synaptic weights are the transpose of the top-down synaptic weights in this model, although this assumption can be relaxed using an approach similar to the one used in variational autoencoders (VAEs) (see “Predictive Coding and the Free Energy Principle” section). The neural response vector $\vct{r}$ is updated based on a weighted combination of the bottom-up prediction error $(\vct{I} - f(\vct{Ur}))$ and the top-down prediction error $\vct{r}^{td} - \vct{r})$ (\cref{fig:hpc}B). Each error is weighted by the inverse of the corresponding noise variance:
The larger the noise variance, the smaller the weight given to that error term, consistent with the concept of Kalman filtering (see section “Prediction in Time: Spatiotemporal Predictive Coding and Kalman Filtering”).

The learning rule for the feedback synaptic weights $\mat{U}$ (and feedforward weights $\mat{U}^\top$) is obtained by using gradient descent on $H$ with respect to $\mat{U}$:
\begin{align}
    \frac{\dd \mathbf{U}}{\dd t}=-\frac{k_2}{2} \frac{\partial H}{\partial \mathbf{U}}=\frac{k_2}{\sigma^2} \frac{\partial f^{\top}}{\partial \mathbf{x}}(\mathbf{I}-f(\mathbf{U} \mathbf{r})) \mathbf{r}^{\top}-k_2 \lambda \mathbf{U}, \label{eq:synap}
\end{align}
where $k_2$ is a positive parameter determining the learning rate of the network and $\vct{x} = \vct{Ur}$. Note that this learning rule is a form of Hebbian plasticity: for the feedforward weights $\vct{U}^\top$, the input presynaptic activity is the residual error $(\vct{I} - f(\vct{Ur}))$ (weighted by $\frac{\dd f^\top}{\dd\vct{x}}$) and the output
postsynaptic activity is $\vct{r}$. More importantly, unlike backpropagation, the learning rule above is local since the feedforward connection explicitly conveys the prediction error at each level. To ensure stability, learning of synaptic weights operates on a slower time scale than the dynamics of $\vct{r}$: The learning rate $k_2$ is a much smaller value than the rate $k_1$ governing the dynamics of the network. For static inputs, this implies that the network responses $\vct{r}$ converge to an estimate for the current input before the synaptic weights $\mat{U}$ are updated based on this converged estimate. An example two-level hierarchical network is depicted in \cref{fig:hpc}C.

\subsubsection{Feedforward Perception as the Initial Inference Step in Predictive Coding}

How does the traditional feedforward “bucket brigade” model of perception, where inputs are processed sequentially in one area and passed on to the next (e.g., LGN \textrightarrow V1 \textrightarrow V2 $\dots$), align with the hierarchical predictive coding view of cortical processing? The answer to this question is easy to obtain from \cref{eq:dynamics} by considering what happens in the very first time step $t=1$ when $\hat{\vct{r}}_0 = \vct{0}$ and the two top-down prediction terms $f(\mat{U}\hat{\vct{r}}_0$ and $\vct{r}^{td}$ are also both $\vct{0}$. In this case, if $g'(\vct{0})$ is also $\vct{0}$, \cref{eq:dynamics} reduces to: 
\begin{align}
    \hat{\mathbf{r}}_t=\frac{k_1}{\sigma^2} \mathbf{U}^{\top} \frac{\partial f^{\top}}{\partial \mathbf{x}_{t-1}} \mathbf{I}.
\end{align}
Thus, the first feedforward pass through the network multiplies the input $\vct{I}$ with the feedforward weights $\mat{U}^\top$ (besides the other multiplicative factors). Assuming this happens at all patches of an image, this equation describes exactly the type of operation implemented by a standard feedforward layer where the filters are given by the rows of $\mat{U}^\top$. In the other words, for a static input, if the top-down predictions are assumed to be zero, a hierarchical predictive coding network (e.g., \cref{fig:hpc}C) initializes its estimates at all levels in the same manner as a deep neural network via a feedforward pass through all layers, before proceeding to further minimize prediction errors by generating top-down predictions from these initial estimates and refining them based on prediction errors.

\subsection{Prediction in Time: Spatiotemporal Predictive Coding and Kalman Filtering}
The model described thus far focused on learning and predicting static inputs. But the world is dynamic -- most of the time, animals receive time-varying stimuli either due to their own movement or due to other moving objects in the environment. This makes the ability to predict future stimuli essential for survival (e.g., predicting the location of predators). The predictive coding framework can be extended to include temporal predictions \parencite{rao_development_1998, rao_optimal_1999}. Specifically, the network dynamics derived above for predictive coding implements a nonlinear and hierarchical form of Bayesian inference that can be related to the classic technique of Kalman filtering \parencite{kalman_new_1960}. This relationship becomes clear when we augment the spatial generative model in \cref{eq:img-gen} with the ability to model the temporal dynamics of hidden state $\vct{r}$ from time step $t$ to $t+1$:
\begin{align}
    \vct{r}_{t+1} = \mat{V}_t\vct{r}_t + \vct{m}_t, \label{eq:time-gen}
\end{align}
where $\mat{V}_t$ is a (potentially time-varying) transition matrix and $\vct{m}_t$ is zero mean Gaussian noise. \cref{eq:time-gen} models how a hidden state in the world, for example, the location of a predator, changes over time by assuming that the next state depends only on the current state (“Markov” assumption) plus some noise. Making the weights $\mat{V}_t$ time-varying allows the equation to capture nonlinear transition dynamics.

Combining \cref{eq:img-gen} with \cref{eq:time-gen} and assuming the function $f$ is the identity function, one can derive the following equations for the network dynamics:
\begin{align}
    \text{Prediction: } & \bar{\vct{r}}_t = \mat{V}_t\vct{r}_{t-1} \nonumber\\
    \text{Correction: } & \hat{\vct{r}}_t = \bar{\vct{r}}_t + \mat{N}_t\mat{U}^\top\mat{G}_t(\vct{I}_t - \mat{U}\bar{\vct{r}}_t), \label{eq:kalman}
\end{align}
where $\mat{N}_t$ and $\mat{G}_t$ are gain terms that depend on the (co-)variances of $\mat{m}$ in \cref{eq:time-gen} and $\vct{n}$ in \cref{eq:img-gen} (see \textcite{rao_optimal_1999} for the derivation). The prediction equation takes the current estimate of the state and generates a prediction of the next state $\bar{\vct{r}}_t$ via the matrix $\mat{V}_t$. The correction equation corrects this prediction $\bar{\vct{r}}_t$ by adding to it the prediction error $(\vct{I}_t - \mat{U}\bar{\vct{r}}_t)$ weighted by gain terms $\mat{N}_t$ and $\mat{G}_t$, with the matrix $\mat{U}^\top$ translating the error from the image space back to the more abstract state space of $\vct{r}$. The gain terms $\mat{N}_t$ and $\mat{G}_t$ could potentially depend on task-dependent factors and can be regarded as “attentional modulation” of the prediction error (see section “Attention and Robust Predictive Coding”) \parencite{rao_correlates_1998}. The above equations implement a Kalman filter (see \textcite{rao_optimal_1999}).

\begin{figure}
    \centering
    \includegraphics[width=\textwidth]{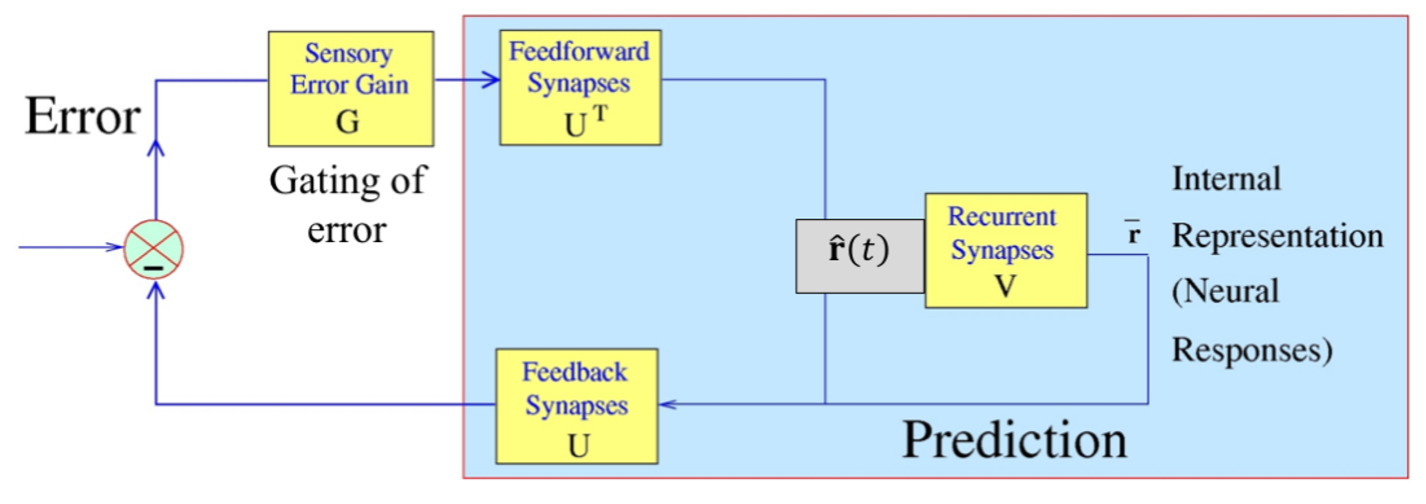}
    \caption{\textbf{Spatiotemporal predictive coding model.} The network is similar to the predictive coding model in \cref{fig:hpc} in terms of the feedback and feedforward pathways conveying prediction and prediction error, respectively, but the spatiotemporal model additionally utilizes local recurrent synapses (lateral connections) to generate the prediction for the next time step. Based on \textcite{rao_optimal_1999}.}
    \label{fig:kalman}
\end{figure}

\cref{fig:kalman} illustrates a neural network implementing the spatiotemporal predictive coding model given by \cref{eq:kalman}: the network uses local recurrent (lateral) connections $\mat{V}$ to make a prediction $\bar{\vct{r}}_t$ for the next time step, translates the prediction to the lower level as $\mat{U}\bar{\vct{r}}_t$ via feedback connections, conveys the prediction error $(\vct{I}_t - \mat{U}\bar{\vct{r}}_t)$ via feedforward connections, and then corrects its state prediction $\bar{\vct{r}}_t$ with prediction error weighted by the gain term $\mat{G}$.

\subsubsection{Prediction and Internal Simulation in the Absence of Inputs}
The spatiotemporal predictive coding model allows for the possibility that the organism or agent might want to perform internal simulations of the dynamics of the external world (e.g., for planning) by predicting how future states evolve given a starting state (and possibly actions). This can be done by setting the input prediction error gain term $\mat{G}_t$ in \cref{eq:kalman} to zero (see also the relationship to attention below). This results in the following network dynamics for a single-level network:
\begin{align}
    \bar{\vct{r}}_t &= \mat{V}_t\hat{\vct{r}}_{t-1} \\ 
    \hat{\vct{r}}_t &= \bar{\vct{r}}_{t}.
\end{align}
In this case, the network ignores any inputs and simply predicts future states moving forward in time using the learned state transition dynamics $\mat{V}_t$. The network thus acts as a recurrent network, with a possibly time-varying set of recurrent weights $\mat{V}_t$ to model nonlinear transitions.

For a hierarchical network, the network dynamics becomes (based on \cref{eq:dynamics}):
\begin{align}
    \hat{\mathbf{r}}_t=\overline{\mathbf{r}}_t+\alpha\left(\mathbf{r}_t^{t d}-\overline{\mathbf{r}}_t\right),
\end{align}
where $\alpha$ is the weight assigned to the prediction $\vct{r}_t^{td}$ from the higher level. Here, the network combines a local recurrent prediction $\bar{\vct{r}}_t$ at one level with a prediction $\vct{r}_t^{td}$ from a higher level (using the weights $(1 - \alpha)$ and $\alpha$ respectively), allowing higher levels to guide the predictions at the lower levels during internal simulation, while ignoring external inputs.

\subsubsection{Attention and Robust Predictive Coding}
The Rao-Ballard predictive coding model can be extended to model top-down attention using a robust optimization function as first proposed in \textcite{rao_correlates_1998}. Specifically, instead of using the squared error loss function
\begin{align}
    H_1 = \frac{1}{\sigma^2}\Vert\vct{I} - f(\mat{U}\vct{r})\Vert_2^2 = \frac{1}{\sigma^2}(\vct{I} - f(\mat{U}\vct{r}))^\top(\vct{I} - f(\mat{U}\vct{r})), 
\end{align}
the robust predictive coding model uses
\begin{align}
    H_1^R = \rho(\vct{I} - f(\vct{Ur})),
\end{align}
where $\rho$ is a function that reduces the influence of outliers (large prediction errors) in the estimation of $\vct{r}$. As an example, $\rho$ could be defined in terms of a diagonal matrix $\mat{S}$ as follows \parencite{rao_correlates_1998}:
\begin{align}
    H^R = \frac{1}{\sigma^2}(\vct{I} - f(\mat{U}\vct{r}))^\top\mat{S}(\vct{I} - f(\mat{U}\vct{r})), 
\end{align}
where the diagonal entries $S_{ii}$ determine the weight accorded to the prediction error at input location $i$: $(I_i - f(\vct{u}_i\vct{r}))^2$ where $\vct{u}_i$ denotes the $i$th row of $\mat{U}$ ($\vct{u}_i$ here is a row vector). A simple but attractive choice for these weights is the nonlinear function given by:
\begin{align}
    S_{ii} = \min\left\{1, \frac{c}{(I_i - f(\vct{u}_i\vct{r}))^2}\right\},
\end{align}
where $c$ is a threshold parameter. This function has the following desirable effect: $\mat{S}$ clips the squared prediction error for the $i$th input location to a constant value $c$ if $(I_i - f(\vct{u}_i\vct{r}))^2$ exceeds the threshold $c$.

Minimizing the robust optimization function $H^R$ leads to the following equation for robust predictive coding:
\begin{align}
    \hat{\mathbf{r}}_t=\hat{\mathbf{r}}_{t-1}+k_1 \mathbf{U}^{\top} \mat{G}_t \frac{\dd f^{\top}}{\dd \mathbf{x}_{t-1}}\left(\mathbf{I} - f\left(\mathbf{U} \hat{\mathbf{r}}_{t-1}\right)\right),
\end{align}
where $\mat{G}_t$ is a diagonal matrix whose diagonal entries at time constant $t$ are given by $G_{ii} = 0$ if $(I_i - f(\vct{u}_i\hat{\vct{r}}_{t-1}))^2 > c_t$ and $1$ otherwise. Here, $c_t$ is a potentially time-varying threshold on the squared prediction error.

\begin{figure}
    \centering
    \includegraphics[width=\textwidth]{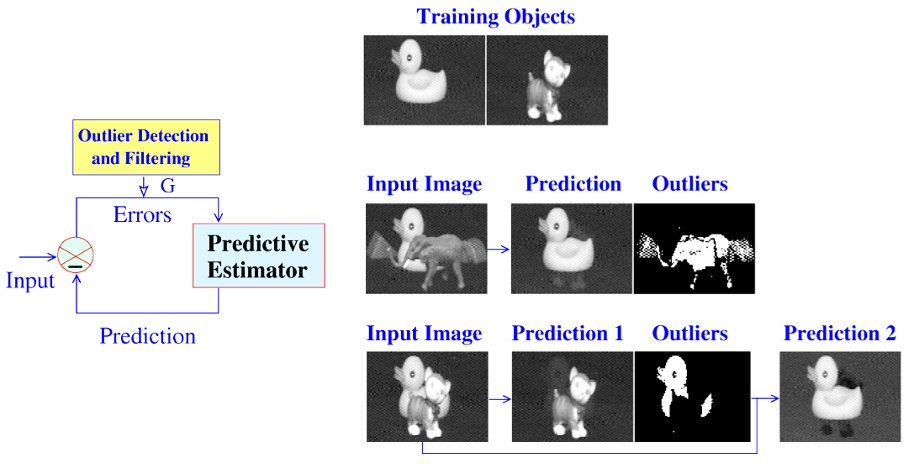}
    \caption{\textbf{Robust predictive coding and attention.} Left panel: The robust predictive coding model utilizes a gain or gating term to modulate the prediction errors before they are fed to the predictive estimator network that estimates the state vector $\vct{r}$. This allows the network to filter out any outliers dynamically as a function of the current top-down hypothesis, allowing the network to “focus attention” and test its hypothesis. Right panel: The top row shows images of two objects the network was trained on. The middle row illustrates how the network can filter out occluders and background objects as outliers, recovering an estimate of a training object (the duck). The bottom row shows how, when presented with an image containing both training objects, the network can sequentially focus and recognize each object.}
    \label{fig:attention}
\end{figure}

The gain $\mat{G}_t$ acts as an “attentional filter” for outlier detection and filtering, allowing the predictive coding network estimating $\vct{r}$ (\cref{fig:attention}, left panel) to suppress large prediction errors in parts of the input containing outliers. This enables the network to focus on verifying the feasibility of its current best hypothesis by trying to minimize prediction errors while ignoring outliers. Robust predictive coding thus allows the network to “focus its attention” on one object while ignoring occluders and background objects, and even “switch attention” from one object to another (\cref{fig:attention}, right panel) (see \textcite{rao_correlates_1998, rao_optimal_1999}).

\subsubsection{What-Where Predictive Coding Networks and Equivariance}
The predictive coding models above do not consider the fact that many natural inputs, such as videos, are generated by the same object or feature undergoing specific transformations such as translations, rotations, and scaling. The predictive coding model has been extended to account for such transformations using “What-Where” predictive coding \parencite{rao_development_1998} and related models that learn transformations based on Lie groups \parencite{rao_learning_1998, miao_learning_2007} and bilinear models \parencite{grimes_bilinear_2005}.

\begin{figure}
    \centering
    \includegraphics[width=\textwidth]{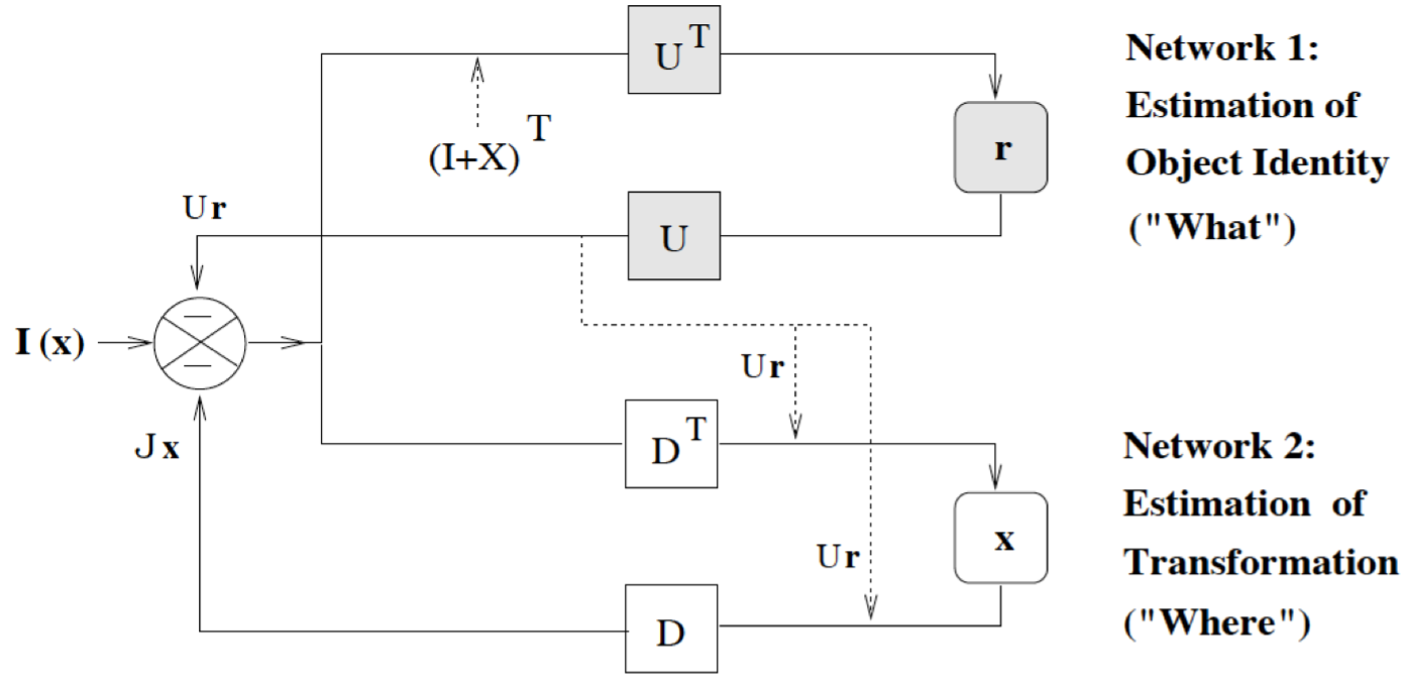}
    \caption{\textbf{What-Where predictive coding networks.} Two networks jointly minimize prediction error: one network estimates object features/identity (“What”) while the other estimates transformations (“Where”) relative to a canonical representation.}
    \label{fig:lie}
\end{figure}

The What-Where predictive coding model is shown in \cref{fig:lie}. It employs two networks to explain a new input $\vct{I}(\vct{x})$: one network, called the “What” network, is similar to the original predictive coding network discussed above and estimates the features or object present in the image via the state vector $\vct{r}$; the other network, called the “Where” network, estimates the transformation $\vct{x}$ in the new input relative to a previous (canonical) input $\vct{I}(\vct{0})$. The network architecture and the dynamics of how $\vct{r}$ and $\vct{x}$ are updated are both derived from first principles through prediction error minimization \parencite{rao_learning_1998}.

The What-Where predictive coding network was one of the first neural networks to demonstrate \textit{equivariance}: the representation of an object in the “What” network remains stable and invariant by virtue of having a second network, the “Where” network, which absorbs changes in the input stream by modeling these changes as transformations of a canonical representation \parencite{rao_development_1998} (cf. the more recent line of research on equivariance using “capsule” networks \parencite{hinton_transforming_2011, sabour_dynamic_2017, kosiorek_stacked_2019}. The What-Where predictive coding model contrasts with traditional deep neural networks which utilize pooling in successive layers to achieve invariance to transformations but at the cost of losing information about the transformations themselves. 

While its architecture is derived from the principle of prediction error minimization, the What-Where predictive coding model shares similarities with the ventral-dorsal visual processing pathways in the primate visual cortex, where ventral cortical areas have been implicated in object-related processing (“What”) and dorsal cortical areas have been implicated in motion- and spatial-transformation-related processing (“Where”). 

\subsubsection{Predictive Coding and the Free Energy Principle}
Predictive coding and the principle of prediction error minimization are closely related to variational inference and learning, which form the basis for VAEs in machine learning research \parencite{dayan_helmholtz_1995, kingma_auto-encoding_2014} as well as the free energy principle in neuroscience as proposed by Friston and colleagues \parencite{friston_theory_2005, friston_free-energy_2010, friston_predictive_2009}. This relationship is briefly summarized below.

MAP inference, as employed in the predictive coding model above, finds an estimate that maximizes the posterior distribution $p(\vct{r} | \vct{I})$. Variational inference aims to find the full posterior distribution instead of a point estimate. Applying Bayes’ rule:
\begin{align}
    p(\mathbf{r} | \mathbf{I})=\frac{p(\mathbf{I} | \mathbf{r}) p(\mathbf{r})}{p(\mathbf{I})}=\frac{p(\mathbf{I} | \mathbf{r}) p(\mathbf{r})}{\int \dd \mathbf{r} p(\mathbf{I} | \mathbf{r}) p(\mathbf{r})}.
\end{align}

The normalizing factor (denominator) contains multidimensional integrals that are usually intractable to compute (e.g., if $p(\vct{r})$ is a sparsity-inducing Laplace distribution in sparse coding). Due to this intractability, variational inference approximates the posterior as follows: the true posterior probability distribution $p_\theta$ parameterized by parameters $\theta$ is approximated with a more tractable distribution $q_\varphi$ parameterized by parameters $\varphi$. The “error” between the two distributions is quantified using the Kullback-Leibler (KL) divergence between the posterior probabilities of the latent variable $\vct{r}$ given the input data $\vct{I}$:
\begin{align}
KL\left(q_{\varphi}(\mathbf{r} | \mathbf{I}) \| p_\theta(\mathbf{r} | \mathbf{I})\right) & =\int q_{\varphi}(\mathbf{r} | \mathbf{I}) \log \frac{q_{\varphi}(\mathbf{r} | \mathbf{I})}{p_\theta(\mathbf{r} | \mathbf{I})} \dd \mathbf{r} \nonumber\\
& =\int q_{\varphi}(\mathbf{r} | \mathbf{I}) \log \frac{q_{\varphi}(\mathbf{r} | \mathbf{I})}{p_\theta(\mathbf{r}, \mathbf{I})} \dd \mathbf{r}+\int q_{\varphi}(\mathbf{r} | \mathbf{I}) \log p_\theta(\mathbf{I}) \dd \mathbf{r} \nonumber\\
& =\int q_{\varphi}(\mathbf{r} | \mathbf{I}) \log \frac{q_{\varphi}(\mathbf{r} | \mathbf{I})}{p_\theta(\mathbf{r}, \mathbf{I})} \dd \mathbf{r}+\log p_\theta(\mathbf{I}) \nonumber\\
& =\mathcal{F}+\log p_\theta(\mathbf{I}), \label{eq:free_en}
\end{align}
where $\mathcal{F}$ is called the “variational free energy” and $\log p_\theta(\vct{I})$ is called the data log likelihood (given model parameters $\theta$) or model evidence. Note that variational free energy $\mathcal{F}$ should not be
confused with the physical notion of free energy (e.g., in thermodynamics), although there is a similarity in their definitions.

Rewriting \cref{eq:free_en}, we have:
\begin{align}
\log p_\theta(\mathbf{I}) & = KL\left(q_{\varphi}(\mathbf{r} | \mathbf{I}) \| p_\theta(\mathbf{r} | \mathbf{I})\right)-\mathcal{F} \nonumber\\
& =KL\left(q_{\varphi}(\mathbf{r} | \mathbf{I}) \| p_\theta(\mathbf{r} | \mathbf{I})\right)+\mathcal{L}, 
\end{align}
where $\mathcal{L} = -\mathcal{F}$ is called the evidence lower bound (or ELBO) in the variational learning and VAE literature since $\log p_\theta(\vct{I}) \geq \mathcal{L}$ (the KL divergence is nonnegative). It can be seen that an organism or artificial agent can increase model evidence (data log likelihood) by maximizing the ELBO $\mathcal{L}$ or equivalently, minimizing variational free energy $\mathcal{F}$ with respect to the latent state and parameters. Note that since $\mathcal{F} = KL\left(q_{\varphi}(\mathbf{r} | \mathbf{I}) \| p_\theta(\mathbf{r} | \mathbf{I})\right) - \log p_\theta(\vct{I})$ and $\log p_\theta(\vct{I})$ does not depend on $\vct{r}$ or $\varphi$, maximizing the ELBO (minimizing $\mathcal{F}$) with respect to $\vct{r}$ and $\varphi$ is equivalent to minimizing the KL divergence between the approximating tractable distribution $q$ and the true distribution $p$. 

To make the connection to predictive coding, the definition of variational free energy $\mathcal{F}$ used in \cref{eq:free_en} can be rewritten as follows:
\begin{align}
    \mathcal{F} &= \int q_{\varphi}(\mathbf{r} | \mathbf{I}) \log \frac{q_{\varphi}(\mathbf{r} | \mathbf{I})}{p_\theta(\vct{I} | \mathbf{r}) p_\theta(\mathbf{r})} \dd \mathbf{r} \nonumber\\
                &= \int q_{\varphi}(\mathbf{r} | \mathbf{I}) \log \frac{q_{\varphi}(\mathbf{r} | \mathbf{I})}{p_\theta(\mathbf{r})} \dd \mathbf{r} + \int q_{\varphi}(\mathbf{r} | \mathbf{I}) \log \frac{1}{p_\theta(\vct{I} | \mathbf{r})} \dd \mathbf{r} \nonumber\\
                &= KL(q_\varphi(\vct{r} | \vct{I}) || p_\theta(\vct{r})) - \mathbb{E}_{q}\left[\log p_\theta(\vct{I} | \vct{r})\right].
\end{align}
Using the relationship in \cref{eq:likelihood} for the negative logarithm of $p_\theta(\vct{I} | \vct{r})$ and using $\alpha$ as the constant of proportionality for \cref{eq:likelihood}, the free energy for the predictive coding model is given by:
\begin{align}
    \mathcal{F} &= KL(q_\varphi(\vct{r} | \vct{I}) || p_\theta(\vct{r})) - \mathbb{E}_{q}\left[-\alpha H_1\right] \nonumber\\
                &= KL(q_\varphi(\vct{r} | \vct{I}) || p_\theta(\vct{r})) + \alpha\mathbb{E}_{q}\left[\frac{1}{\sigma^2}\Vert\vct{I} - f(\vct{Ur})\Vert_2^2\right] \nonumber\\
                &= \text{KL divergence between posterior and prior for } \vct{r} + \alpha\times \text{mean squared prediction error}.
\end{align}
Thus, within the predictive coding framework, minimizing the variational free energy $\mathcal{F}$, as advocated by the free energy principle of brain function \parencite{bogacz_tutorial_2017, friston_free-energy_2010}, is equivalent to finding an approximating posterior distribution $q_\varphi$ that both minimizes prediction errors while also attempting to be close to the prior for $\vct{r}$. This can be regarded as a full-distribution version of the predictive coding model described above, which uses MAP inference to find an optimal point estimate that minimizes prediction errors while also being constrained by the negative logarithm of the prior (\cref{eq:loss}).

\subsubsection{Action-Based Predictive Coding and Active Inference}
Prediction error can be minimized not only by estimating optimal hidden states $\vct{r}$ (perception) and learning optimal synaptic weights $\mat{U}$ and $\mat{V}$ (internal model learning) but also by choosing appropriate actions. Inferring actions that minimize prediction error with respect to a goal, or more generally, a prior distribution over future states, is called \textit{active inference} \parencite{fountas_deep_2020, friston_action_2011, friston_active_2017}. For example, in a navigation task, if the objective is to reach a desired goal location by passing through a series of landmarks, prediction error with respect to the goal and landmarks can be minimized by selecting actions at each time step that reach each landmark and eventually the goal location. Active inference can be regarded as an example of “planning by inference” where an internal model is used to perform Bayesian inference of actions that maximize expected reward or the probability of reaching a goal state \parencite{attias_planning_2003, botvinick_planning_2012, verma_goal-based_2005, verma_planning_2006}.

Predictive coding allows internal models for action inference to be learned by predicting the sensory consequences of an executed action. For example, babies, even in the womb, make seemingly random movements called “body babbling” \parencite{nehaniv_bayesian_2007} that can allow a predictive coding network to learn a mapping between the current action and the sensory input received immediately after. After learning such an action-based prediction model via prediction error minimization, the model can be unrolled in time into the future to specify a desired goal state (or states) (see, e.g., \textcite{verma_goal-based_2005, verma_planning_2006}), and predictive coding-based inference can used to infer a set of current and future actions most likely to lead to the goal state(s). Some of the empirical evidence reviewed in the section “Empirical Evidence for Predictive Coding” on visual and auditory predictions based on motor activity can be understood within the framework of action-based predictive coding.

\section{Predictive Coding in the Visual System}

\subsection{Predictive Coding in Early Stages of Visual Processing}
Early “predictive coding” models focused on explaining the center–surround response properties and biphasic temporal antagonism of cells in the retina \parencite{atick_could_1992, buchsbaum_trichromacy_1983, meister_neural_1999, srinivasan_predictive_1982} 
and lateral geniculate nucleus (LGN) \parencite{dong_statistics_1995, dan_efficient_1996}. These models were derived from the information-theoretic principle of efficient coding \parencite{attneave_informational_1954, barlow_possible_1961, } rather than hierarchical generative models like the Rao-Ballard model. Under the efficient coding hypothesis, the goal of the visual system is to efficiently represent visual information by reducing redundancy arising from natural scene statistics \parencite{dong_temporal_1995, field_relations_1987, ruderman_statistics_1994}. A simple example of redundancy reduction is to remove aspects of an input that are predictable from nearby inputs. Neural activities then only need to represent information that deviates from the prediction.

\textcite{srinivasan_predictive_1982} proposed that the spatial and temporal receptive field properties of retinal ganglion cells are a result of predicting local intensity values in natural images from a linear weighted sum of nearby values in space or preceding input values in time. Training a linear system that predicts the pixel intensity at a location from its surrounding pixels produces prediction weights that closely resemble the receptive fields of retinal ganglion cells \parencite{huang_predictive_2011, srinivasan_predictive_1982}. Thus, the neural activities of retinal ganglion cells can be seen as representing the “whitened” residual errors that the system cannot predict. Srinivasan et al. also showed that the linear predictor weights depend on the signal-to-noise (SNR) ratios of visual scenes. Larger groups of neighboring regions need to be integrated in order to cancel out high statistical noise in low SNR input, a phenomenon observed by the authors in the fly eye. More recently, \textcite{hosoya_dynamic_2005} showed that retinal ganglion cells can rapidly adapt to environments with changing correlation structure and become more sensitive to novel stimuli, consistent with the predictive coding view of the retina.

Similar ideas have been used to cast LGN processing as performing temporal whitening of inputs from the retina \parencite{atick_could_1992, dan_efficient_1996, dong_statistics_1995, kaplan_information_1993}. \textcite{dong_statistics_1995} derived a linear model whose objective is to produce decorrelated output in the frequency domain. The optimized spatiotemporal filter compares remarkably well with the physiological data from the LGN \parencite{saul_spatial_1990}. \textcite{dan_efficient_1996} confirmed through experiments that the output from the LGN is temporally decorrelated (especially for lower-frequency 3–15 Hz) for natural stimuli but not white noise, suggesting that the LGN selectively whitens stimuli that match natural scene statistics. In summary, these results suggest that the early stages of visual processing (the retina and LGN) are tuned to the statistical properties of the natural environment. The same insight, implemented via a hierarchical generative model, forms the core of the Rao-Ballard predictive coding model of the visual cortex.

\subsection{Predictive Coding in the Visual Cortex}
\begin{figure}
    \centering
    \includegraphics[width=\textwidth]{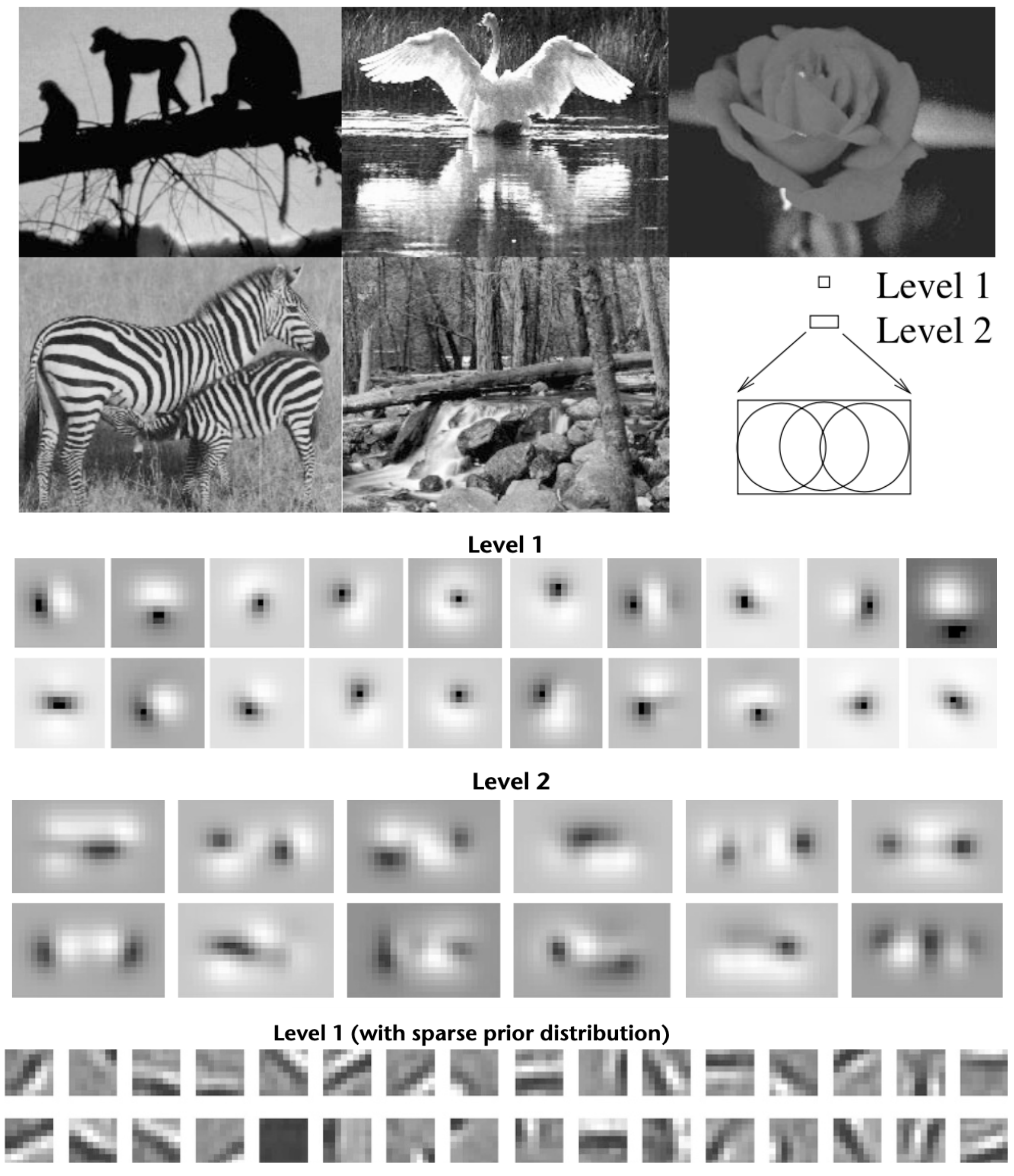}
    \caption{\textbf{Emergence of visual cortex–like receptive fields in the hierarchical predictive coding model.} Top panel: Natural images used for training the hierarchical model. Several thousand natural image patches were extracted from these five images. The bottom-right corner shows the size of Level 1 and Level 2 receptive fields relative to the natural images. Middle two panels: Level 1 and Level 2 feedforward synaptic weights (rows of $\mat{U}^\top$) learned from the natural images using a Gaussian prior. Values can be zero (always represented by the same gray level), negative (inhibitory, black regions), and positive (excitatory, bright regions). These weights resemble centered oriented Gabor-like receptive fields in Level 1 and in Level 2, various combinations of the synaptic weights in Level 1. Bottom panel: Localized Gabor filter-like synaptic weights learned in Level 1 using a sigmoidal nonlinear generative model and a sparse kurtotic prior distribution similar to those used in \textcite{olshausen_emergence_1996}. Adapted from \textcite{rao_predictive_1999}.}
    \label{fig:hpc_result}
\end{figure}

The model presented in “Hierarchical Predictive Coding” was used by Rao and Ballard to explain both classical and extra-classical receptive fields effects in the visual cortex in terms of prediction error minimization. The cortex is modeled as a hierarchical network in which higher-level neurons predict the neural activities of lower-level neurons via feedback connections (\cref{fig:hpc}A, lower arrows). A class of lower-level neurons, known as “error neurons,” compute the differences between the predictions from the higher level and the actual responses at the lower level, and convey these prediction errors back to the higher level via feedforward connections (\cref{fig:hpc}A, upper arrows). Except for neurons at the highest level, neural activities at every level are influenced by both “top-down” predictions and “bottom-up” prediction errors (\cref{fig:hpc}B). Additionally, the network is structured such that the higher-level neurons make predictions at a larger spatial scale than lower-level neurons; this is achieved by allowing higher-level neurons to predict the responses of several lower-level modules, resulting in a combined receptive field larger than any single lower-level neuron’s receptive field (e.g., in \cref{fig:hpc}C, a single Level 2 module predicts the responses of three Level 1 modules).

The dynamics of the recurrent neural network implementing predictive coding is governed by \cref{eq:dyn} and the synaptic weights are learned using \cref{eq:synap}. When trained on natural image patches (\cref{fig:hpc_result}, top panel), the synaptic weights that were learned in the first level resembled oriented spatial filters or Gabor wavelets similar to the receptive fields of simple cells in V1 while at the second level, the synaptic weights resembled more complex features that appear to be combinations of several lower-level filters (\cref{fig:hpc_result}, Level 2).

\subsection{Endstopping and Contextual Effects as Prediction Error Minimization}
\begin{figure}
    \centering
    \includegraphics[width=0.8\textwidth]{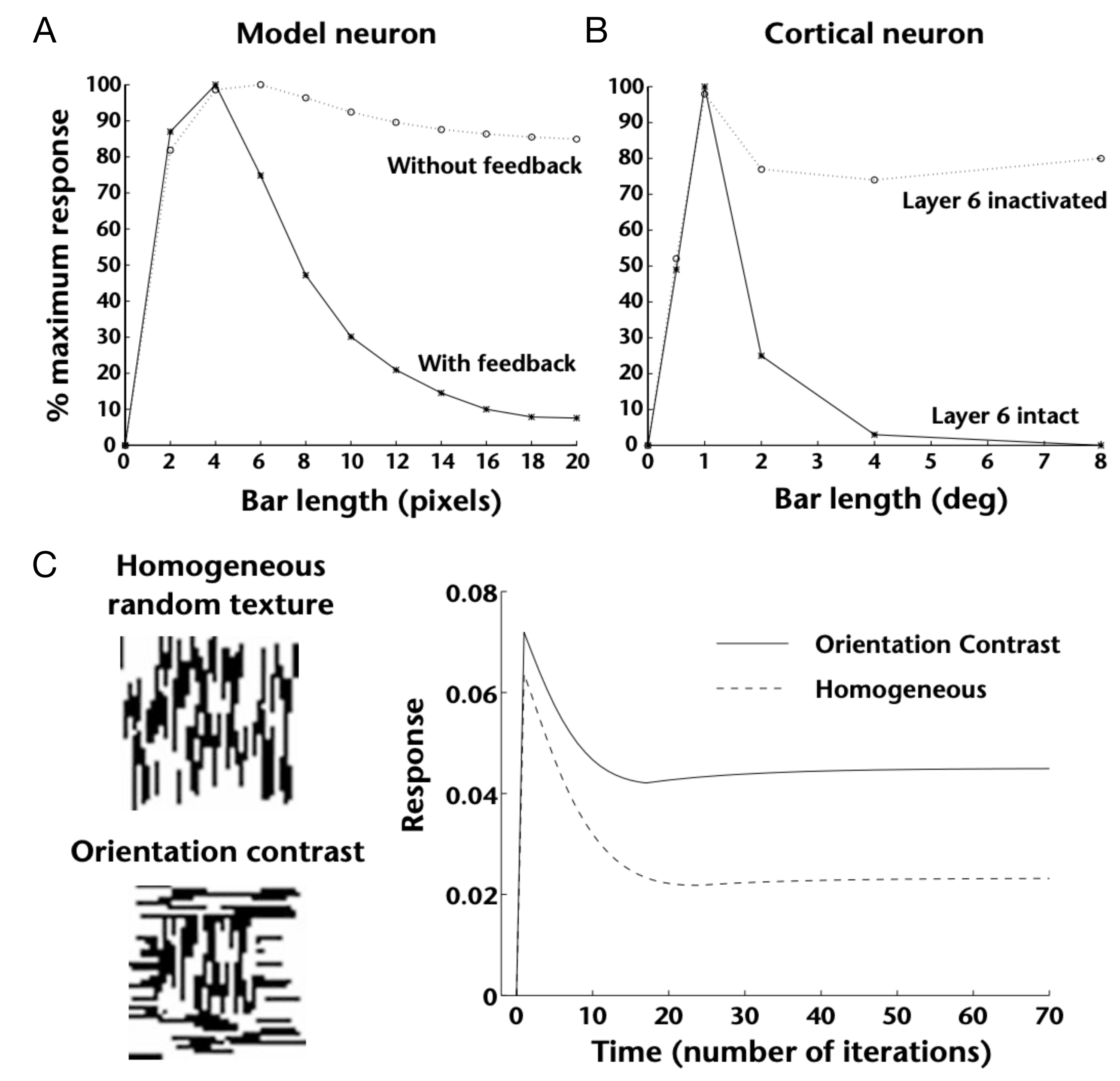}
    \caption{\textbf{Extra-classical receptive field effects in the hierarchical predictive coding model.} \textbf{(A)} Tuning curve of a first level model neuron for the control case (solid curve) and after ablation of feedback from the second level (dotted curve). \textbf{(B)} Endstopping in a layer 2/3 complex cell in cat primary visual cortex. Tuning curves with respect to bar stimulus length are shown for the control case (solid curve) and after inactivation of layer 6 (dotted curve). \textbf{(C)} Contextual modulation effects. Responses of an error-detecting model neuron for oriented texture stimuli with center and surround regions having the same (dotted line) versus different (solid line) orientations. Adapted from \textcite{rao_predictive_1999}}
    \label{fig:endstopping}
\end{figure}

Some visual cortical neurons (particularly those in layers 2/3) exhibit the curious property that a strong response to a stimulus gets suppressed when a stimulus is introduced in the surrounding region whose properties (e.g., orientation) match the properties of the stimulus at the center of the receptive field (RF). Such effects, which have been reported in several cortical areas \parencite{bolz_generation_1986, desimone_visual_1987, hubel_receptive_1968}, are known as “extra-classical” receptive field effects or contextual modulation. Hubel and Wiesel named one class of such cells in area V1 “hypercomplex” cells and noted that these cells exhibit the property of “endstopping”: The cell’s response is inhibited or eliminated when an oriented bar stimulus in the center of the cell’s RF is extended beyond its RF to the surrounding region.

\textcite{rao_predictive_1999} suggested that endstopping and related contextual effects could be interpreted in terms of prediction errors in a network trained for predictive coding of natural images. The responses of neurons representing prediction errors (e.g., neurons in cortical layers 2/3 that send axons to a “higher” cortical area) are suppressed when the top-down prediction becomes more accurate because the larger stimulus (e.g., longer bar) engages higher-level neurons tuned to this stimulus. These neurons generate more accurate predictions for the lower level, resulting in low prediction errors. When the surrounding context is missing or at odds with the central stimulus, the prediction error responses are high due to the mismatch between the higher level’s prediction and the lower-level responses. Rao and Ballard proposed that the tendency for the higher level to expect similar statistics (e.g., similar orientation) for a central patch and its surrounding region arises from the statistics of natural images that exhibit such statistical regularities and the fact that the hierarchical predictive coding network has been trained as a generative model to emulate these statistics.

\cref{fig:endstopping} illustrates the prediction error responses from a two-level predictive coding network trained on natural images. The error-detecting model neurons at the first level (with firing rates $\vct{r} - \vct{r}^{td}$) display endstopping similar to cortical neurons (\cref{fig:endstopping}B, solid curve): Model neuron responses are suppressed when the bar extends beyond the classical receptive field (figure 7A, solid curve) as the predictions from the higher level become progressively more accurate with longer bars. Elimination of predictive feedback causes the error-detecting neurons to continue to respond robustly to longer bars (\cref{fig:endstopping}A, dotted curve). The same model can also explain contextual effects (\cref{fig:endstopping}C): The first-level error detecting neurons show greater responses (solid line) when the texture stimulus at the center has the same orientation as the stimulus in the surround compared to an orthogonally oriented surround stimulus (dashed line). Similar contextual effects have been reported in V1 neurons \parencite{zipser_contextual_1996}. Other V1 response properties such as cross-orientation suppression and orientation contrast facilitation can also be explained by the predictive coding framework \parencite{spratling_reconciling_2008, spratling_predictive_2010}.

In summary, the predictive coding model suggests that (a) the physiological properties of visual cortical neurons are a consequence of statistical learning of an internal model of the natural environment—specifically, the objective of prediction error minimization allows the cortex to learn a hierarchical generative model of the natural world; and (b) perception is the process of actively explaining input stimuli by inverting a learned internal generative model via inference to recover hidden causes of the input. Context effects such as endstopping arise as a natural consequence of the visual cortex detecting prediction errors or deviations from the expectations generated by a learned internal model of the natural environment.

\subsection{A Common Misconception About the Predictive Coding Model}
\begin{figure}
    \centering
    \includegraphics[width=\textwidth]{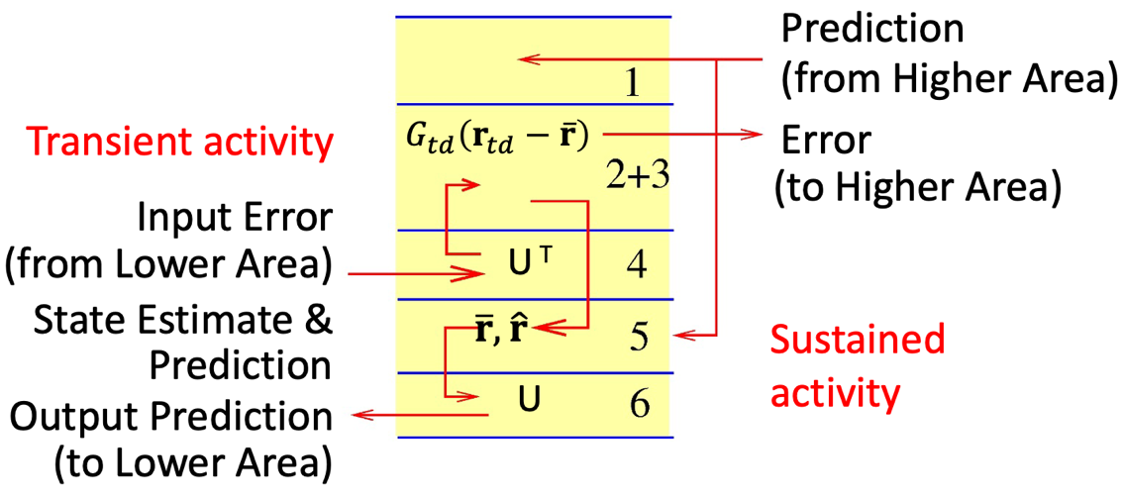}
    \caption{\textbf{Putative laminar implementation of the Rao-Ballard predictive coding model.} A mapping of the hierarchical predictive coding model components (see \cref{fig:hpc,,fig:kalman}) to cortical layers.}
    \label{fig:laminar}
\end{figure}

One of the most common misconceptions about the predictive coding model is that the model predicts suppression of all neural activity when stimuli become predictable. This has led some authors to state that experimental evidence showing neurons not being suppressed or maintaining persistent firing for predictable inputs contradicts the predictive coding model. On the contrary, the predictive coding model requires a group of neurons to maintain the internal representation (state estimate $\hat{\vct{r}}$) at each hierarchical level for generating predictions for the lower level (see “Hierarchical Predictive Coding” and \cref{fig:laminar}). Thus, in the predictive coding model, the neurons that are suppressed when stimuli become predictable are error-detecting neurons that are distinct from the neurons maintaining the network’s internal representation of the external world. Similar to the efficient coding models of the retina and LGN \parencite{srinivasan_predictive_1982, dong_statistics_1995}, redundancy reduction occurs primarily in the feedforward pathways of the Rao-Ballard predictive coding model, with the feedback pathways remaining active to convey predictions.

\subsection{Neuroanatomical Implementation of Predictive Coding}
\textcite{rao_predictive_1999} postulated two groups of neurons at each hierarchical level with distinct computational goals (\cref{fig:laminar}). One group of neurons maintains an internal representation (state estimate) for generating top-down predictions of lower-level activities. These neurons are hypothesized to be in the deep layers 5/6 of cortical columns and are predicted by the model to exhibit sustained activity to maintain predictions to lower levels. A different group of neurons at the same level calculates prediction errors to be conveyed to the next higher level. These were suggested to be layer 2/3 neurons which send connections to “higher” order cortical areas and which are expected to exhibit transient activity. Since prediction errors can be positive or negative, \textcite{rao_predictive_1999} proposed two subclasses of error-detecting neurons, one subclass representing positive errors and another representing negative errors, similar to on-center off-surround and off-center on-surround neurons in the retina and LGN.

In general, as seen above in endstopping and other contextual effects, the model predicts that layer 2/3 neurons are suppressed when the stimuli are predictable (i.e., consistent with natural image statistics) while deeper layer neurons remain active. Stimuli that deviate from natural image statistics (“novel” stimuli) on the other hand elicit large responses in layer 2/3 neurons. The model also predicts that prediction error signals are used for unsupervised learning of the synaptic connections in the predictive coding network, driving the synaptic weights to better reflect the structure of the input stimuli.

\section{Empirical Evidence for Predictive Coding}
Experimental evidence has been mounting for predictive processing in the cortex thanks to advances in neuronal recording and stimulation techniques such as optical imaging and optogenetics. Particularly relevant to the hierarchical predictive coding model proposed by \textcite{rao_predictive_1999} are findings of top-down predictive “internal representation” neurons and bottom-up error-detecting neurons in a cortical column. These findings appear to suggest that the cortex may indeed be implementing a hierarchical generative model of the natural world. We briefly review the experimental evidence below.

\subsection{Internal Representation Neurons and Prediction Error Neurons in the Cortex}
The hierarchical predictive coding model predicts the existence of at least two functionally distinct classes of neurons in the cortex: internal state representation neurons $\vct{r}$, which maintain the current estimate of state at a given hierarchical level and are postulated to reside in the deeper layers 5/6 of the cortex, and error-detecting neurons $\vct{r} - \vct{r}^{td}$ in layers 2/3, which compute the difference between the current state estimate and its top-down prediction from a higher level. Recent studies have provided evidence for both types of neurons in the cortex.

\textcite{keller_sensorimotor_2012} recorded neural activities from layer 2/3 cells in the monocular visual cortex of behaving mice that were head-fixed and running on a spherical treadmill. The mice were exposed to 10–30 minutes of visual feedback as they ran on the treadmill. In normal “feedback” trials, the visual flow stimuli provided to the mouse were full-field vertical gratings coupled to the mouse’s locomotion on the treadmill. In “mismatch” trials, visual-locomotion mismatches were delivered randomly as brief visual flow halts (1 second). As a control, the mice also went through “playback trials” in which visual flow was passively viewed without locomotion.

\begin{center}
    \includegraphics[width=\textwidth]{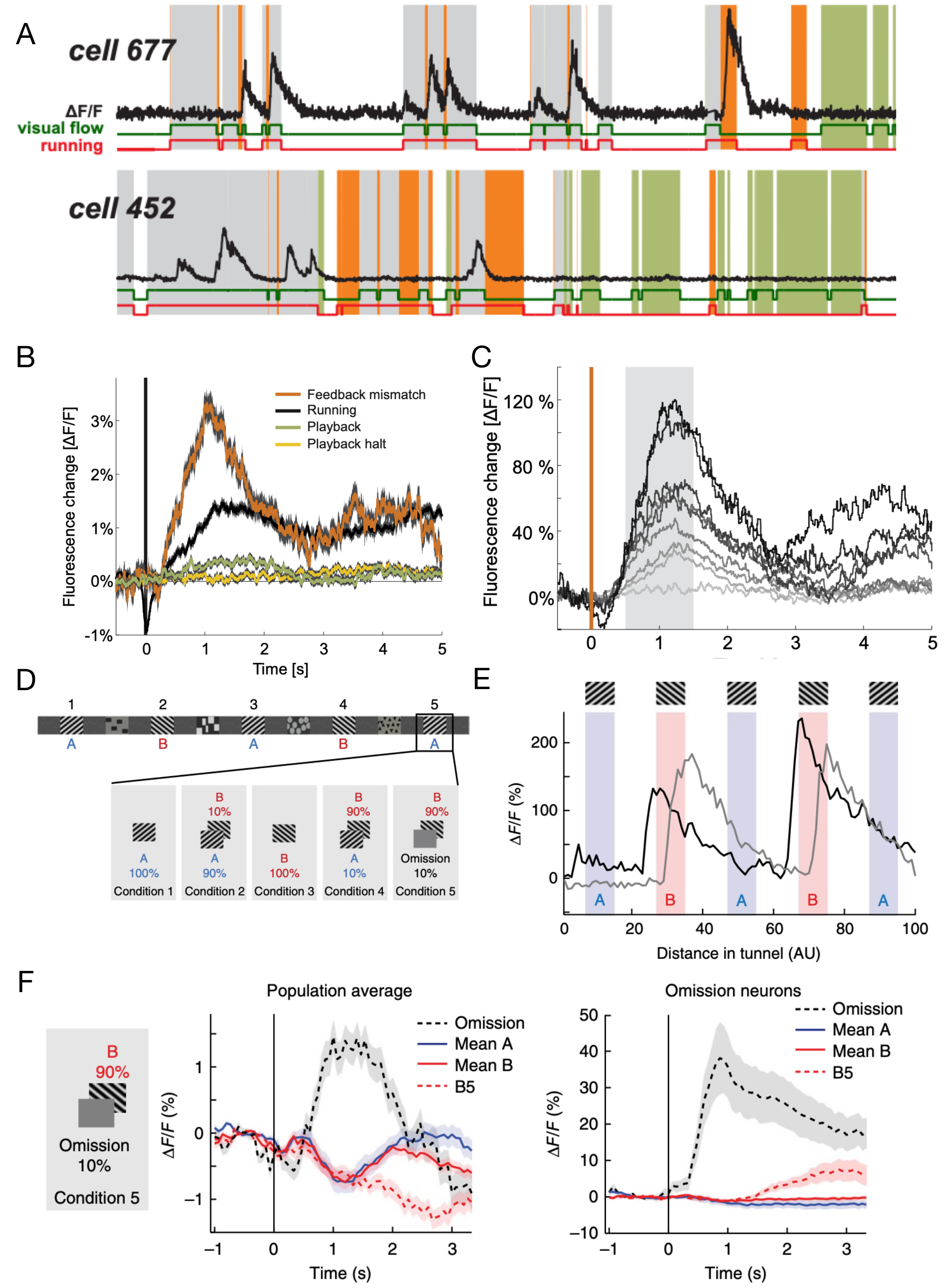}
    \captionof{figure}{\textbf{Evidence for predictive neurons and error-detecting neurons in two visual-locomotion tasks.} \textbf{(A)} Sample fluorescence change ($\Delta F/F$) traces from two cells (black). Green trace: binary indicator of visual flow. Red trace: binary indicator of locomotion. Gray shading: visual feedback with locomotion. Orange shading: feedback mismatch (no visual flow with motion). Green shading: playback (visual flow with no motion). White: baseline. Cell 677 responded predominately to feedback mismatch while cell 452 responded mainly to visual feedback with locomotion. \textbf{(B)} Average population response to onsets (time=0) of different trials: feedback mismatch (orange), running (black), playback (green), and passive viewing of playback halts (yellow). \textbf{(C)} Mismatch responses during various speeds of locomotion. Darker traces represent faster locomotion speed. \textbf{(D)} Schematic representation of the texture lining both walls of the virtual tunnel that mice ran through. The last block indicates the percentage of appearance of each texture (including omission trials) under different conditions. \textbf{(E)} Two example neurons’ response traces when the mouse ran through the tunnel. Blue shading represents the period when the mouse was in Block A, red shading represents Block B. Both neurons are spatially selective only to Block B, not Block A. The neuron depicted by the black trace showed predictive responses (activation prior to entering Block B), while the neuron depicted by the gray trace was only responsive after the mouse entered Block B. \textbf{(F)} left: average population response to omission trials (black dashed) compared to Block A trials (blue), Block B trials (red), and the 90\% Block B5 trials (red dashed). Panel F, right: average responses of the omission-selective neural population. Adapted from \textcite{keller_sensorimotor_2012} (Panels A–C) and \textcite{fiser_experience-dependent_2016} (Panels D–F).}
    \label{fig:keller}
\end{center}

The authors found that 13.0\% of the visual cortical neurons recorded responded predominately to feedback mismatches. \cref{fig:keller}A shows a sample neuron (cell number 677) that responded mainly to mismatch trials (orange shading). Also, 23.6\% of the neurons responded mainly to feedback trials in which visual flow feedback was predictable (cell number 452 in \cref{fig:keller}A). The mismatch responses were also significant in the population average (\cref{fig:keller}B) and the activity onset in mismatch trials was much stronger than that in the other trials. Furthermore, the mismatch signals encoded the degree of mismatch -- a visual flow halt during faster locomotion resulted in a stronger response than during slow locomotion (\cref{fig:keller}C, darker lines denote faster speed at the time of visual flow halt).

V1 neurons have also been found to be predictive of spatial locations after adapting to a new environment. In an experiment by \textcite{fiser_experience-dependent_2016}, mice went through a virtual tunnel with blocks of two different grating patterns (A or B) separated by distinct landmarks. The five trial conditions only differed in the fifth block, where the grating patterns A and B as well as omission with no visual stimuli had different probabilities of occurring (see \cref{fig:keller}D). After adaptation, some neurons developed predictive responses to specific visual stimuli based on spatial information. As shown in \cref{fig:keller}E, an example neuron (black trace) showed strong activation before the mouse perceived Block B (but not Block A). In contrast, another sample neuron (gray trace) showed activation after entering Block B (but not Block A). The authors also discovered prediction error responses similar to those reported by \textcite{keller_sensorimotor_2012}. The population average of neural activities during omission trials was much greater than during A and B trials (\cref{fig:keller}F, left). Moreover, a subset of neurons (2.3\%) developed omission-selectivity -- they showed large responses only to the omission trials (\cref{fig:keller}F, right).

Other studies have also documented neural responses carrying predictive information. \textcite{xu_activity_2012} found that after rats adapted to a visual moving dot trajectory, a brief flash at the starting point of the same trajectory triggered the same sequential firing pattern in the rat’s V1 as evoked by the full-sequence stimulus. Similarly, \textcite{gavornik_learned_2014} discovered that after an animal is exposed to a sequence of stimuli during training, V1 regenerates the sequential response even when certain elements of the sequence are omitted. 

Prediction and prediction error-like signals have also been found in cortical areas in the human visual cortex (e.g., \textcite{murray_shape_2002}) and the hierarchical face processing region of the monkey inferior temporal cortex (IT) \parencite{tsao_cortical_2006, freiwald_functional_2010}. \textcite{schwiedrzik_high-level_2017} exposed macaque monkeys to fixed pairs of face images with different head orientations and identities such that the successor face image can be predicted from the preceding face image. Neurons in the lower-level face area ML (middle lateral section of the superior temporal sulcus) displayed large responses when the pair association was violated (either in identity, or head orientation, or both). Furthermore, prediction errors resulting from view violation (head orientation) diminished and eventually vanished during the late phase of responses while those resulting from identity violation remained significant. This is consistent with the interpretation that the top-down predictive signals from the view-invariant neurons in higher-level anterior lateral and anterior medial areas suppress the view mismatch responses (encoded locally in the lower-level ML area), while identity-related mismatch signals are propagated through feedforward circuits for further processing. In another study, \textcite{issa_neural_2018} used different face-part configuration stimuli (typical versus atypical) and found that the lower-level areas of the hierarchy (posterior IT and central IT) signal deviations of their preferred features from the expected configurations, whereas the top level (anterior IT) maintained a preference for natural, frontal face-part configuration. The authors further discovered that the early responses in central IT and anterior IT are correlated with late responses in posterior IT: Images that produced large responses in higher-level areas early are followed by reduced activities in lower-level areas, consistent with top-down predictions signal subduing lower-level responses.

In another experiment, \textcite{choi_predictive_2018} showed that a hierarchical inference model could explain the effect of feedback signals from the prefrontal cortex to intermediate visual cortex V4 as top-down predictions of partially occluded shapes.

\textcite{schneider_cortical_2018} explored the effects of learning on prediction error-like activity in the primary auditory cortex. Rats were given artificial auditory feedback coupled to their locomotion: The pitch of the sound was proportional to the rat’s running speed. They found that a group of neurons in the rat’s primary auditory cortex initially responded strongly to the artificial auditory feedback (“reafferent sound”) but over the course of several days, the neuronal circuits learned to suppress this activity. The suppression occurred whenever the reafferent sound was coupled to the rat’s locomotion and did not occur when a nonreafferent sound was played or when the reafferent sound was played during resting. The gradual suppression of responses is consistent with how the predictive coding model learns an internal model of the environment: as the network learns to predict the artificial sound coupled to the rat’s locomotion, the predictions get better, resulting in decreasing prediction errors which manifest as suppression of the auditory neurons’ activities.

The results discussed above provide evidence for predictive neural activity and prediction error-like responses in the cortex. The Rao and Ballard model additionally postulates that layer 2/3 neurons compute and convey the prediction errors while neurons in the deeper layers 5/6 maintain the state estimate. Recent experiments have attempted to test these predictions. While it is hard to distinguish the state estimating “internal representation” neurons from those driven by bottom-up sensory stimuli (see review by \textcite{keller_predictive_2018} for further discussions), there is a growing body of evidence suggesting that layer 2/3 neurons may indeed play a role in comparing bottom-up information and top-down predictions.

\subsection{Layer 2/3 Neurons as Top-Down Bottom-Up Signal Comparators}
For biological networks to use prediction errors to correct their estimate, both positive and negative errors need to be represented. At any input location, a positive prediction error ($(\vct{I} - \vct{Ur} > 0)$) occurs when the input is not predicted (or incorrectly predicted) while a negative prediction error ($(\vct{I} - \vct{Ur} < 0)$) occurs when a predicted input is omitted. \textcite{rao_predictive_1999} postulated that layer 2/3 in the cortex may employ two different groups of neurons, one to convey positive errors and another for negative errors, similar to on-center, off-surround and off-center, on-surround ganglion cells in the retina \parencite{srinivasan_predictive_1982}.

To test this theory, \textcite{jordan_opposing_2020} used an experimental setup similar to the one used in \textcite{keller_sensorimotor_2012}: mice ran on a treadmill with locomotion-coupled visual flow feedback. Whole-cell recordings were obtained from both layer 2/3 and layer 5/6 neurons in V1. Visual feedback could be interrupted with a brief flow halt (1 second) at random times to generate visual-locomotion mismatch events. Out of 32 neurons recorded in layer 2/3, 17 neurons showed depolarizing activities (\cref{fig:jordan}A, left, depolarizing mismatch (dMM) neurons) and 6 neurons showed hyperpolarizing activities (\cref{fig:jordan}A, right, hyperpolarizing mismatch (hMM) neurons) during mismatch trials. These results suggest that the dMM and hMM neurons in layer 2/3 may subserve the function of encoding positive and negative prediction errors.

In addition, 30\% of the neurons exhibited significant correlations between the mismatch responses and the speed of locomotion (visual halts that occurred during faster locomotion generated “stronger” mismatch signals). The sign of the correlation was also different between dMM and hMM neurons, with dMM neurons showing a positive correlation (\cref{fig:jordan}B, left) and hMM neurons showing a negative correlation (\cref{fig:jordan}B, right). These results are consistent with Keller and colleagues’ calcium imaging study previously discussed \textcite{keller_sensorimotor_2012} (\cref{fig:keller}C), showing that the responses of layer 2/3 neurons could potentially signal the quantitative level of prediction errors.

\begin{center}
    \includegraphics[width=0.9\textwidth]{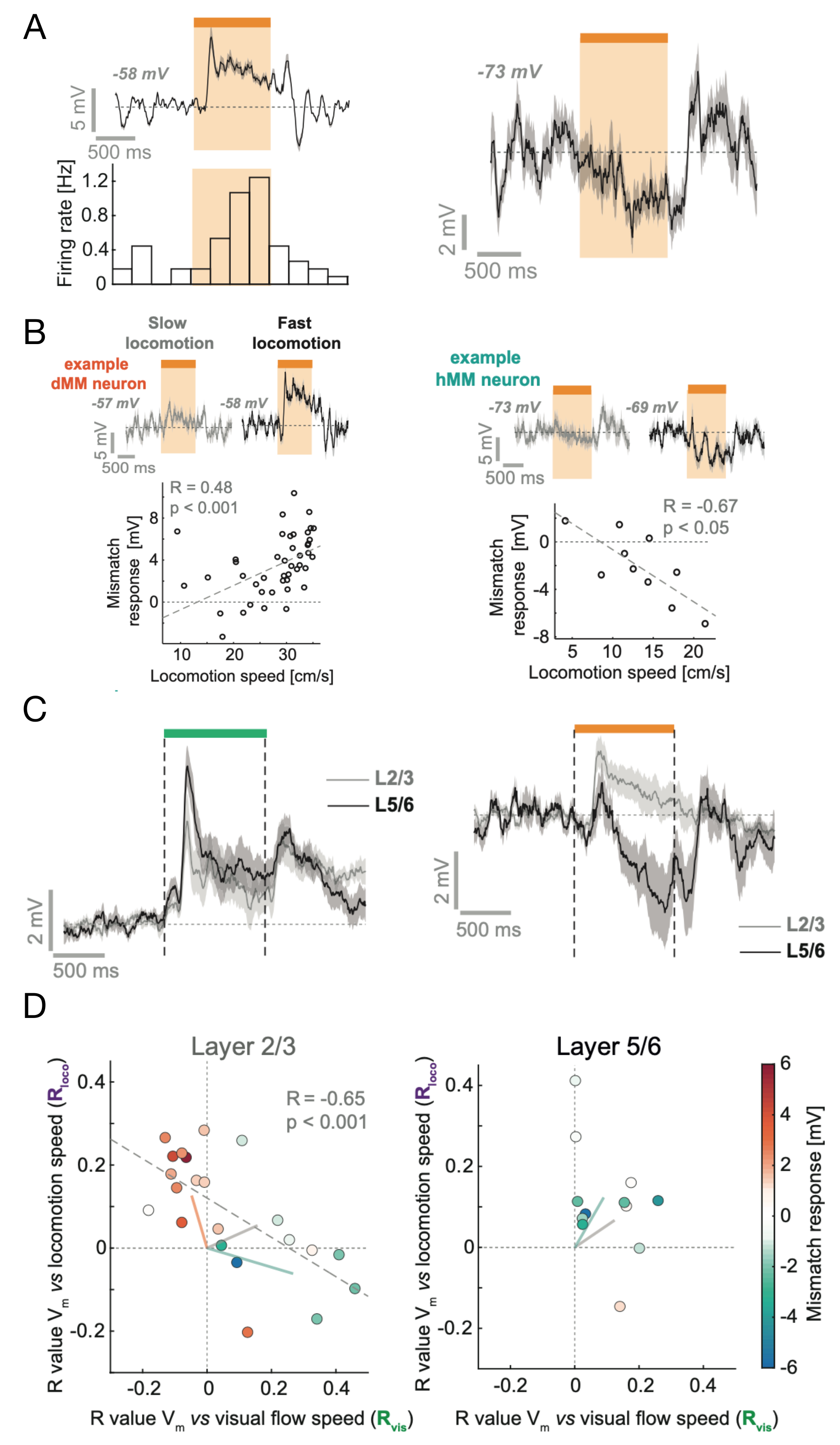}
    \captionof{figure}{\textbf{Comparison of layer 2/3 versus layer 5/6 neurons in the mouse visual cortex.} \textbf{(A)} left: Average membrane potential (Vm) response (top) and firing rate histogram (bottom) during mismatch trials of a layer 2/3 neuron (dMM), showing depolarizing Vm. Orange shading represents the mismatch trial period. The mean prestimulus voltage is –58mV. Panel A, right: Average Vm of another layer 2/3 neuron (hMM) showing hyperpolarizing Vm during mismatch trials. \textbf{(B)}, left: An example dMM neuron showing a positive correlation between response and locomotion speed. Panel B, right: An example hMM neuron showing a negative correlation between response and locomotion speed. \textbf{(C)} Comparison of average responses between layer 2/3 neurons (light gray) and layer 5/6 neurons (dark gray) during visual feedback trials (left) and during mismatch trials (right). \textbf{(D)} Scatter plot of correlation coefficients between the membrane potential of the neurons and visual flow speed (x-axis) or locomotion speed (y-axis). The dot color denotes the membrane potential change in response in mismatch trials. The gray dashed line is the fitted linear regression line to the data. Solid lines represent the average responses of dMM neurons (red), hMM neurons (turquoise), and unclassified neurons (gray). Panel D shows layer 2/3 neurons (left) and layer 5/6 neurons (right). Adapted from \textcite{jordan_opposing_2020}.}
    \label{fig:jordan}
\end{center}

\textcite{jordan_opposing_2020} also investigated the differences between the responses of layer 2/3 neurons and deeper layer 5/6 neurons during normal visual feedback trials and mismatch trials. A much lower ratio of neurons in layers 5/6 (5 out of 14) responded predominately to mismatch trials. Additionally, larger activity during mismatch trials was rare (1 neuron), with 7 neurons exhibiting reduced activities. The difference in responses between superficial and deep layer neurons was significant in mismatch trials (\cref{fig:jordan}C, right) but not in normal visual flow trials (\cref{fig:jordan}C, left). To further characterize the influence of visual flow and locomotion on layer 2/3 neurons versus layer 5/6 neurons, correlations between the activities of these neurons and locomotion speed or visual flow speed were calculated. As seen in \cref{fig:jordan}D (left plot), the distribution of correlations in layer 2/3 was bimodal: Activities of most dMM neurons were positively correlated with locomotion speed and negatively correlated with visual flow speed (and vice versa for hMM neurons). On the other hand, activities of layer 5/6 neurons were mostly positively correlated with both locomotion speed and visual flow speed (\cref{fig:jordan}D, right). These results suggest that layer 2/3 neurons are well-suited to computing the error between the locomotion-generated predictions of visual inputs and the actual visual input, whereas the deeper layer 5/6 neurons may integrate top-down predictions (here, from motor areas) and bottom-up input to compute an estimate of the state at the current hierarchical level.

The difference in neural responses to expected versus unexpected visual flows in layer 2/3 versus layer 5 was also confirmed in a recent study by \textcite{gillon_learning_2021}. The authors used an open-loop experiment (no sensorimotor coupling to locomotion) with stimuli consisting of moving squares. Expectation violations were created in some trials by making 25\% of the visual squares move in the opposite direction compared to the other 75\%. The authors found that somatic and distal apical dendritic populations in layer 5 did not exhibit significantly different responses to expected versus unexpected visual flow, whereas both layer 2/3 somatic and distal apical dendritic populations showed a significant difference in responses. Additionally, this difference increased over days of exposure. Gillon et al. also found learning effects when mice were exposed to Gabor sequence stimuli for several consecutive days. The responses to unexpected stimuli (in this case, novel Gabor stimuli replacing an expected stimulus in a sequence) were predictive of how these responses evolve in subsequent sessions on a cell-by-cell basis. Besides implicating layer 2/3 neuron in prediction error computation, these results further confirm that the neural responses to unexpected stimuli (i.e., prediction errors) can drive learning in neural circuits, an important computational prediction of the predictive coding model \parencite{rao_predictive_1999} (see \cref{eq:synap}).

The larger distribution of error detecting neurons in superficial layers than deep layers was also confirmed by \cite{hamm_cortical_2021} in awake mice with visual oddball paradigms. The authors additionally showed that optogenetic suppression of prefrontal inputs to V1 reduced the contextual selectivity of the error detecting neurons, consistent with the effect of top-down signals in the predictive coding model. Finally, through laminar local field potential recordings in monkeys, \textcite{bastos_layer_2020} showed that predictability of visual stimuli affects neural activities in the superficial and deep layers differently—during predictable trials, there was an enhancement of alpha and beta power in the deep layers of the cortex whereas during unpredictable trials, an increase in spiking and gamma power was observed in the superficial layers.

\section{Discussion}
By casting Bayesian inference and learning in terms of minimizing prediction errors based on an internal model of the world, predictive coding provides a unifying view of perception and learning. Perception is equated with Bayesian inference of hidden states of the world and proceeds by forming predictive hypotheses about inputs that are corrected based on prediction errors. Learning corresponds to using the inferred states to build an internal model of the world that minimizes prediction errors through synaptic plasticity. Actions can further minimize prediction errors with respect to future goals via active inference.

The hierarchical predictive coding model \parencite{rao_predictive_1999} assumes that the hierarchical structure of the cortex forms predictive hypotheses at multiple levels of abstractions to explain input data. The model postulates that feedback connections between cortical areas convey predictions of expected neural activity from higher to lower levels, while the feedforward connections convey the prediction errors back to the higher level to correct the neural activity at that level, characteristics that differentiate hierarchical predictive coding from other cortical models \parencite{lee_hierarchical_2003, heeger_theory_2017}.

Early empirical support for the hierarchical predictive coding model was based on its ability to explain extra-classical receptive field effects such as endstopping and other contextual modulation of responses in the visual cortex in terms of prediction error minimization \parencite{rao_predictive_1999}. Rao and Ballard proposed that neurons in layer 2/3 exhibiting such effects can be interpreted as error-detecting neurons whose responses are suppressed when the properties of stimuli in the center of the receptive field can be predicted by stimuli in the surround, following natural image statistics. Several recent experimental studies have discovered neurons in the visual and auditory cortex that encode predictions or prediction errors in a variety of sensory-motor tasks \parencite{keller_sensorimotor_2012, fiser_experience-dependent_2016, schneider_cortical_2018}. Some studies have tested more detailed neuroanatomical predictions such as the role of cortical layer 2/3 neurons in error computation \parencite{jordan_opposing_2020}. Others have shown that these error-related neural activities can drive learning in synaptic connections \parencite{gillon_learning_2021}. Although further tests are required, the experimental results reviewed above support the hypothesis that the cortex implements a predictive model of the world, uses this model to generate predictions, and utilizes prediction errors to both correct its moment-to-moment estimates and to learn a better model of the world.

There remain many aspects of predictive coding that require further exploration and experimental corroboration. For example, are layer 5/6 neurons computing and maintaining the hidden state as specified by \cref{eq:dyn}? Are the inverse variances in \cref{eq:dyn} (“precisions” terms in the free energy principle; see \textcite{friston_free-energy_2010}) computed in the cortex? If so, how are they used to weigh the bottom-up and top-down terms in the predictive coding network dynamics (\cref{eq:dyn})? How is this “precision”-based weighting related to attention and robust predictive coding \parencite{rao_correlates_1998, rao_optimal_1999}? More broadly, can “what-where” predictive coding networks be made hierarchical and be used to understand visual processing in the ventral and dorsal streams of the visual cortex?

Spatiotemporal hierarchical predictive coding is another area worthy of further study. \textcite{palmer_predictive_2015} derived a model by solving the information bottleneck problem \parencite{tishby_information_2000} and suggested that retinal ganglion cells may signal predictive information about the future states of the environment, a result recently confirmed by \textcite{liu_learning_2021}. \textcite{rao_optimal_1999} presented a single-level Kalman filtering model for predicting inputs one time-step ahead based on learning linear transition dynamics from input sequences. These models, however, do not address hierarchical representation of temporal information. Experimental evidence suggests that cortical representations exhibit a hierarchy of timescales from lower-order to higher-order areas across both sensory and cognitive regions \parencite{murray_hierarchy_2014, runyan_distinct_2017, siegle_survey_2021}. Recent work by the authors \parencite{jiang_dynamic_2021} suggests that a hierarchical predictive coding model based on dynamic synaptic connections (via “hypernetworks”) can learn visual cortical space-time receptive fields and hierarchical temporal representations from natural video sequences. Ongoing work is focused on exploring the connections between such learned temporal representations and response properties in different cortical areas.

The original predictive coding model of Rao and Ballard described how a hierarchical network can converge to \textit{maximum a posteriori} estimates of hidden states at different hierarchical levels. Although the model included variances for the top-down and bottom-up errors, it did not explicitly represent uncertainty. The Kalman filter version of predictive coding \parencite{rao_optimal_1999} does represent uncertainty in terms of a Gaussian posterior distribution, but whether the cortex can compute covariance matrices (or just the diagonal variances) remains unclear. Other theories of how the brain may represent uncertainty and perform Bayesian inference using population coding and sampling \parencite{echeveste_cortical-like_2020, huang_bayesian_2016, orban_neural_2016, rao_bayesian_2004, rao_bayesian_2005} are complementary to predictive coding and the connections between these theories remain to be worked out.

Finally, there is much to be explored in relating predictive coding to cognition, memory, and behavior. Several studies have shown that prediction errors (or “surprise”-related signals) can drive memory reactivation and reconsolidation \parencite{bein_mnemonic_2020, kim_pruning_2014, rust_remembering_2021, sinclair_prediction_2019}, suggesting a role for error signals in memory updating, but the connections to predictive coding theories remain unclear. Friston and colleagues have made important contributions in establishing some of these connections \parencite{friston_free-energy_2010, friston_active_2017} through the free energy principle and active inference (see the section “Predictive Coding and the Free Energy Principle”). Empirical studies such as those reviewed above have demonstrated the close links between predictive coding and active behaviors such as locomotion. We expect future predictive coding theories to incorporate actions, attention, memory, and planning. Together with new tools such as Neuropixels probes \parencite{jun_fully_2017, steinmetz_neuropixels_2021} for large-scale recordings and optogenetics for stimulation, predictive coding theories can enable new paradigms for theory-driven experimentation in neuroscience.

\section{Acknowledgement}
This material is based upon work supported by the Defense Advanced Research Projects Agency (contract number HR001120C0021); the National Institute of Mental Health (grant number 5R01MH112166); the National Science Foundation (grant number EEC-1028725); and a grant from the Templeton World Charity Foundation. The opinions expressed in this publication are those of the authors and do not necessarily reflect the views of the funders. The authors would like to thank Ares Fisher, Dimitrios Gklezakos, and Samantha Sun for suggestions, discussions, and manuscript edits.

\printbibliography

\end{document}